\documentclass[manuscript,screen]{acmart}

\usepackage[linesnumbered,lined,ruled,commentsnumbered]{algorithm2e}
\usepackage{threeparttable}
\usepackage{tcolorbox}
\usepackage{booktabs}
\usepackage{multirow} 
\usepackage{amsmath}
\usepackage{cleveref}
\usepackage{colortbl}
\usepackage{graphicx}
\usepackage{paralist}
\usepackage{framed}
\usepackage{xcolor}
\usepackage{xspace}
\usepackage{url}
\usepackage{listings}
\usepackage{xcolor}
\usepackage{multicol}
\usepackage[T1]{fontenc}
\usepackage{fvextra}
\usepackage{longtable}
\usepackage{enumitem}

\DefineVerbatimEnvironment{Highlighting}{Verbatim}{
    commandchars=\\\{\},
    fontsize=\footnotesize,
    formatcom=\bfseries
}

\definecolor{gray1}{rgb}{0.61, 0.75, 0.91}
\definecolor{gray2}{rgb}{0.77, 0.85, 0.95}
\definecolor{gray3}{rgb}{0.88, 0.92, 0.97}
\definecolor{gray4}{rgb}{0.86, 0.91, 0.92}

\newcommand{\ie}{\textit{i}.\textit{e}.,\xspace}
\newcommand{\eg}{\textit{e}.\textit{g}.,\xspace}

\newcommand{\lgc}{LLMgCode\xspace}
\newcommand{\hac}{HaCode\xspace}
\newcommand{\llm}{LLMs\xspace}
\newcommand{\per}{\textsc{Perplexity}\xspace}


\AtBeginDocument{%
  }

\acmJournal{JACM}
\acmVolume{1}
\acmNumber{1}
\acmArticle{1}
\acmMonth{1}

\begin{document}

\title{One Size Does Not Fit All: Investigating Efficacy of Perplexity in Detecting LLM-Generated Code}

\author{Jinwei Xu}
\orcid{0009-0004-5157-1118}
\email{jinwei\_xu@smail.nju.edu.cn}

\author{He Zhang}
\orcid{0000-0002-9159-5331}
\email{hezhang@nju.edu.cn}
\authornotemark[1]

\author{Yanjing Yang}
\orcid{0009-0006-8789-4589}
\email{yj\_yang@smail.nju.edu.cn}

\author{Lanxin Yang}
\orcid{0000-0002-0406-2263}
\email{lxyang@nju.edu.cn}
\authornote{Corresponding Authors: Lanxin Yang and He Zhang.}

\author{Zeru Cheng}
\orcid{0009-0001-5094-4427}
\email{zeru\_cheng@smail.nju.edu.cn}

\author{Jun Lyu}
\orcid{0000-0001-9070-7269}
\email{lvjun\_dnt@outlook.com}

\author{Bohan Liu}
\orcid{0000-0002-0146-5411}
\email{bohanliu@nju.edu.cn}

\author{Xin Zhou}
\orcid{0000-0002-3263-1275}
\email{zhouxin@nju.edu.cn}
\affiliation{%
  \institution{State Key Laboratory of Novel Software Technology, Software Institute, Nanjing University}  
  \city{Nanjing}
  \country{China}
}

\author{Alberto Bacchelli}
\orcid{0000-0003-0193-6823}
\email{bacchelli@ifi.uzh.ch}
\affiliation{%
  \institution{Department of Informatics, University of Zurich}
  \city{Zurich}
  \country{Switzerland}
}

\author{Yin Kia Chiam}
\orcid{0000-0003-1107-7719}
\email{yinkia@um.edu.my}

\author{Thiam Kian Chiew}
\orcid{0000-0002-4272-7860}
\email{tkchiew@um.edu.my}
\affiliation{%
  \institution{Department of Software Engineering, Faculty of Computer Science and Information Technology, University of Malaya}  
  \country{Malaysia}
}

\renewcommand{\shortauthors}{Xu et al.}

\begin{abstract}
Large language model-generated code (\lgc) has become increasingly common in software development. 
So far \lgc has more quality issues than human-authored code (\hac). 
It is common for \lgc to mix with \hac in a code change, while the change is signed by only human developers, without being carefully examined. 
Many automated methods have been proposed to detect \lgc from \hac, in which the perplexity-based method (\textbf{\per} for short) is the state-of-the-art method. 
However, the efficacy evaluation of \per has focused on detection accuracy. 
Yet it is unclear whether \per is good enough in a wider range of realistic evaluation settings. 
To this end, we carry out a family of experiments to compare \per against feature- and pre-training-based methods from three perspectives: \emph{detection accuracy}, \emph{detection speed}, and \emph{generalization capability}.
The experimental results show that \per has the best generalization capability while having limited detection accuracy and detection speed. 
Based on that, we discuss the strengths and limitations of \per, \eg \per is unsuitable for high-level programming languages.
Finally, we provide recommendations to improve \per and apply it in practice.
As the first large-scale investigation on detecting \lgc from \hac, this article provides a wide range of findings for future improvement.  
\end{abstract}

\ccsdesc[500]{Software and its engineering~Software verification and validation}
\ccsdesc[500]{Software and its engineering~Software supply chain security}

\keywords{Code authorship attribution, human-authored code, LLM-generated code, large language model, perplexity}

\maketitle

\section{Introduction}
The advent of Large Language Models (LLMs) has profoundly advanced automated software engineering~\cite{chen2021evaluating}, of which code generation is most popular~\cite{gu2024effectiveness}. 
Code generation is to generate source code based on natural descriptions~\cite{li2023skcoder}.
Various LLMs (\eg GitHub Copilot~\cite{copilot}, GPT-4~\cite{gpt-4}) have been proposed for code generation in recent years~\cite{du2024evaluating}.
LLMs have the strong capability to follow instructions~\cite{chung2024scaling}, enabling developers to write code by providing natural language descriptions of requirements only~\cite{jiang2024survey}.
Leveraging natural language understanding and generative power, LLMs achieve remarkable success in code generation~\cite{austin2021program}.
Therefore, LLM-generated code (\textbf{\lgc}) has become increasingly common in software development~\cite{sun2024ai, yu2024large}.  

Although \lgc help developers reduce manual efforts and focus on high-level tasks~\cite{liu2024exploring}, the quality and security of \lgc are challenging~\cite{pearce2022asleep, asare2023github, fu2023security, liu2024no}. 
According to \citet{pearce2022asleep}, 40.73\% of the 1,689 programs generated by GitHub Copilot are vulnerable, with risks far more than human-authored code (\textbf{\hac})~\cite{perry2024ccsinsecure, sandoval2023lost}. 
Moreover, \lgc has the risks of plagiarism and copyright infringement~\cite{guo2023aigc}, which are less in \hac. According to \citet{guo2024copyleft}, \lgc can be a copy of \hac even if the latter is with copyright protection~\cite{yu2023codeipprompt}. Due to this, GitHub Copilot was accused of compensating \$9 billion for generating restrictive code without attaching the necessary copyright or licenses~\cite{copilot2022}.

In addition to the challenges of \lgc itself, the mixture of \lgc and \hac has two challenges. The first is compromising the transparency of software development, which results in inaccurate attribution of code ownership in bug accountability and obscured productivity assessments~\cite{shi2024between}.
The second is increasing the difficulty of evaluating the reliability of third-party software, which results in distrust for upstream dependencies in the software supply chain~\cite{bukhari2023distinguishing}.
Therefore, It is important to detect \lgc from \hac. 
To this end, many studies~\cite{shi2024between, suh2025empirical} and commercial tools~\cite{copyleak} have proposed to detect \lgc from \hac, which supports mitigating various risks by taking measures such as conducting stricter security inspections and performing code plagiarism detection~\cite{xu2025distinguishing}. 

The past few years have seen many studies~\cite{shi2024between, bukhari2023distinguishing, idialu2024whodunit, nguyen2024gptsniffer, solaiman2019release, gehrmann2019gltr, ippolito2019automatic, mitchell2023detectgpt, su-etal-2023-detectllm} that proposed methods to detect \lgc from \hac. 
According to the detection mechanisms these methods use, they can be divided into three categories: 
\begin{inparaenum}
    \item \emph{feature}-based,
    \item \emph{pre-training}-based, and 
    \item \emph{perplexity}-based methods (\textbf{\per} for short).
\end{inparaenum}
\emph{Feature}-based methods~\cite{bukhari2023distinguishing, idialu2024whodunit} are generally conventional machine learning classifiers that require professionals to devise appropriate features from various aspects such as lexical and syntactic characteristics.
\emph{Pre-training}-based methods~\cite{nguyen2024gptsniffer} take the code snippets as inputs. They convert code snippets to tokens, embed them as vectors, and use the pre-trained model as a classifier. 
\per\footnote{For succinctness, \per refers to the family of perplexity-based methods throughout the paper unless it is specifically stated otherwise.}~\cite{shi2024between, solaiman2019release, gehrmann2019gltr, ippolito2019automatic, mitchell2023detectgpt, su-etal-2023-detectllm} is a class of heuristic methods, which assume that \lgc has a lower perplexity than \hac.
Technically, perplexity is a measure of how confident a probability model is in predicting a sample~\cite{xu2024detecting}. 
When it comes to code generation, perplexity is used to measure the predictability of the next token of code based on prompts and existing tokens. The calculation of the perplexity is based on an LLM. 

\begin{figure}[htbp]
    \begin{center}
    \includegraphics[width=0.99\textwidth]{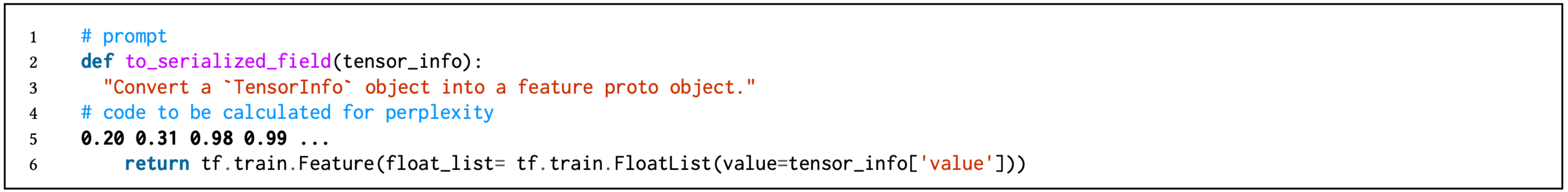}
    \end{center}
    \caption{An example of calculating the perplexity of a set of existing code snippets}
\label{FIG:example-0} 
\end{figure}

Figure 1 presents an example of calculating the perplexity of a probability model (CodeLlama) for code snippets (either \lgc or \hac). 
The prompt is shown in lines 2-3, and the code snippets to be calculated are in line 6. 
Based on the prompt, LLM calculates the probability of ``~~'' (Tab) as the first token is 0.20. Based on the prompt and ``~~'', the LLM calculates the probability of ``\texttt{return}'' as the second token is 0.31. Followed by ``\texttt{~tf}'' and ``\texttt{.train}'' are 0.98 and 0.99, respectively. 
The higher the probability of a token, the more confident the model is about generating this token. Therefore, the perplexity of the model in understanding this token is low. 
Perplexity is a probability-based measure and has many implementations (\eg log of probability~\cite{solaiman2019release}, log of rank~\cite{ippolito2019automatic}, and log of probability divided by log of rank~\cite{su-etal-2023-detectllm}) in which the \emph{log of probability} is common. For example, the perplexity of CodeLlama for ``\texttt{return}'' is $ln(0.31)=-1.18$.
The perplexity of a model for a piece of code is usually the average perplexity of tokens. 

When the code is natural and coherent, the model is confident in predicting the next token based on the prompts and the existing tokens, resulting in a lower perplexity. 
According to the mechanism of \llm for code generation, \llm generate code in the unit of a token through understanding previous tokens, calculating the probability of each possible next token, and applying the sampling strategy to select one with high probability as the output.
Compared with \lgc, \hac is more casual and personalized, making it more unpredictable. 
Therefore, it is reasonable that \lgc has lower perplexity than \hac. 
Due to its explainability and scalability (\eg log of probability, log of rank), \per has been widely used~\cite{solaiman2019release, gehrmann2019gltr, ippolito2019automatic, mitchell2023detectgpt, su-etal-2023-detectllm}.

Although \per is becoming the most popular method, evaluating \per is limited to detection accuracy. 
For example, \citet{shi2024between} evaluated the proposed DetectCodeGPT using AUC (\underline{A}rea \underline{U}nder the \underline{C}urve) on Python only. 
Yet it is unclear what the detection accuracy of DetectCodeGPT is other than Python. 
DetectCodeGPT detects only the code generated by \llm with fewer than 7B parameters; its effectiveness on the commonly used commercial models such as GPT-4o and Gemini-1.0 is unclear.
Moreover, DetectCodeGPT uses the same model for code generation and perplexity calculation. As it is hard to know in advance what model is for code generation, the effectiveness of DetectCodeGPT in a wide range of contexts is unclear. The limited evaluation of \per does not reflect its effects well in practice. 

Toward realistic evaluations of \per, we conduct a large-scale investigation on \per from three perspectives: \textbf{detection accuracy}, \textbf{detection speed}, and \textbf{generalization capability}. 
Detection accuracy is the most fundamental goal of detecting \lgc, while it might not be good between situations, \eg an effective method for C++ may not suit Python. In this article, we evaluate the accuracy in various situations from three aspects. 
Specifically, the detection accuracy (effectiveness) is evaluated from the aspects of the programming language (C, C\#, C++, Go, Java, Python, Ruby, and Rust), degree of difficulty (easy, middle, and hard), and scale of solution (small, middle, and large). 
Efficiency greatly affects the user experience and large-scale applications; thereby, the detection speed (seconds/sample) of \per should be considered.
Since new powerful \llm are emerging, developers utilize various \llm to assist in programming, the generalization capability of \per is important.
The generalization capability is evaluated from the aspect of code generation models (GPT-3.5, GPT-4o, Gemini-1.0, and Llama-3.1).  
We hope the multi-perspective efficacy evaluations can reflect the efficacy of \per in practice. 
With experimental evaluation, we first devised a dataset containing 729 programming puzzles, each accompanied by at least two \hac solutions in each of the eight programming languages available in CodeNet~\cite{puri2021codenet}. To obtain \lgc solutions, we used GPT-3.5 to generate two solutions per puzzle in each language, resulting in a total of 11,664 \lgc solutions. For each specific evaluation aspect, we randomly sampled 500 \lgc and 500 \hac solutions for each category. For example, when it comes to programming languages, we sampled 500 \lgc and 500 \hac solutions for each language.
To evaluate the generalization capability of \per, we generated 500 \lgc solutions using each of GPT-4o, Gemini-1.0, and Llama-3.1, ensuring a consistent sample size with other evaluation aspects. Specifically, we randomly selected 32 puzzles from the dataset and used each LLM (GPT-4o, Gemini-1.0, and Llama-3.1) to generate two solutions per puzzle in each language. Each LLM thus produced 512 \lgc solutions, from which we randomly sampled 500 to form the final set of its solutions. 
Then, we compared eight perplexity-based methods against two feature-based methods and one pre-training-based method that have been used to detect \lgc from \hac. 

The experimental results show that (1) for detection accuracy, the best-performing \per performs well in detecting \lgc in C/C++ (AUC=91\%) and large-scale code (AUC=88\%) but still does not perform as well as the other two methods regarding programming language, degree of difficulty, and scale of solution; (2) for detection speed, \per without perturbation is as efficient as others, while \per with perturbation have a relatively slow efficiency, with detection speed ranging from 10 to 150 seconds per sample; (3) for generalization capability, \per perform better than others across different \llm, with AUC ranging from 80\% to 82\%. 
Based on experimental results, we provide implications of \per from both positive and negative perspectives. On the negative side, \per is unsuitable for high-level programming languages, short code snippets, and just-in-time detection. On the positive side, \per exhibits good generalization capability, which performs well for C/C++, and has good interpretability. Finally, we provide three directions for improving \per in the future and recommendations for the application of \per.

The main contributions of this article can be summarized as follows.
\begin{itemize}
    \item \textbf{Experimental design}: An experimental design to investigate the detection accuracy, detection speed, and generation capability of \per in detecting \lgc.
    \item \textbf{Findings}: A wide range of findings that illustrate the advantages and limitations of \per in detecting \lgc.
    \item \textbf{Dataset}: A large-scale dataset for continuous research on detecting \lgc.
\end{itemize}

The remainder of this article is organized as follows. \Cref{SEC:BAC} presents background and related work. \Cref{SEC:MOT} illustrates motivating examples. \Cref{SEC:MED} and \Cref{SEC:RES} elaborate on the research methodology and results, respectively. \Cref{SEC:IMP} discusses implications. Finally, we present threats to validity in \Cref{SEC:TTV} and conclude this article in \Cref{SEC:CON}.

\section{Background and Related Work}
\label{SEC:BAC}
This section presents an overview of \llm for code generation and reviews studies that apply \per for code authorship attribution.

\subsection{Background: Large Language Models for Code Generation}

Since \llm were proposed, they have been widely applied in many tasks across a variety of domains~\cite{bommasani2021opportunities, wei2022emergent, zhao2023survey, wu2023bloomberggpt}.
One of the most exciting applications of \llm is code generation~\cite{brown2020language, dong2024self, hong2023metagpt, zhang2023planning, zhou2022docprompting}. 
\llm are capable of generating source code based on prompts. 
The prompts consist of natural and programmatic descriptions of code functions, retractions, input and output formats, or any other elements.
Following the prompts, an LLM generates tokens sequentially until it arrives at a stop indicator or the token limit. 
 
Many studies show that \lgc helps developers save time~\cite{coignion2024performance}, reduce effort~\cite{ziegler2022productivity}, and improve productivity~\cite{siddiq2024sallm} in software development. 
Moreover, \lgc help ease developers as they sometimes suffer from heavy workload~\cite{dunne2024weaknesses}. With \lgc, developers can pay more attention to important and urgent tasks.
Because of these benefits, \lgc has been widely used and has even become a must-have in modern software development~\cite{spiess2025calibration, north2024code}. 
A survey made with 500 USA developers showed that 92\% of respondents use \lgc for official assignments and personal use~\cite{shani2023survey}.  

Although \llm bring many benefits, the quality and security of \lgc remain challenging~\cite{pearce2022asleep, yu2024codereval, dakhel2023github}. 
One of the causes is that the corpus used for training \llm is not qualified and secure enough~\cite{asare2023github}. 
Therefore, \lgc might be buggy and vulnerable~\cite{ schuster2021you, jesse2023large}. 
For example, \citet{fu2022potential} assessed the security of \lgc in six common programming languages and found that 156 out of 435 code segments contained security vulnerabilities. 
The proportion of vulnerabilities in C++ and Go code accounts for 46.1\% and 45.0\%, respectively. 
\citet{mastropaolo2023robustness} assessed the robustness of \lgc and found that slight modifications to the natural description while still maintaining semantic equivalence would result in \llm generating different results, occurring in 46\% (408/892) of the total samples. 
\citet{jimenez2023swe} assessed the capability of large language models to solve 2294 real-world problems (Issues) on GitHub and found that the pass rate of code generated using GPT-4 has the lowest pass rate--0\%, using Claude 2 has the highest pass rate--1.96\%.
In addition, \lgc might be similar or even unchanged from existing code snippets in the training corpus~\cite{guo2024copyleft}. 
As the training corpus might include a significant amount of restrictive code~\cite{yu2023codeipprompt}, there are copyright infringement risks for \lgc.

\subsection{Related Work: Perplexity for Code Authorship Attribution}
Detecting \lgc from \hac can be regarded as a code authorship attribution problem~\cite{abuhamad2018large, dauber2019git, bogomolov2021authorship} in which the authors are either \llm or humans.
Code authorship attribution fills in the missing or inaccurate author information of source code fragments, which is beneficial for software maintenance~\cite{girba2005developers, anvik2006should, fritz2010degree} and software quality analysis~\cite{rahman2011ownership, bird2011don, yin2011fixes, thongtanunam2016revisiting}. 
For example, code authorship attribution helps detect code plagiarism~\cite{burrows2007source, kothari2007probabilistic} and trace malware provenance~\cite{caliskan2015anonymizing}. 
Moreover, detecting \lgc from \hac increases the authenticity and integrity of software~\cite{shi2024between}, reducing a variety of risks such as security vulnerabilities and copyright infringement in the software supply chain~\cite{bukhari2023distinguishing}.
In addition, \lgc is typically not considered original~\cite{nguyen2024gptsniffer}, thus detecting \lgc is helpful in scenarios that require individual work, such as programming contests~\cite{idialu2024whodunit,nguyen2024gptsniffer}. This scenario also asks for a high detection efficiency, as examination systems need to provide quick feedback after submission.

\per is originally one of the major methods used for detecting AI-generated content (including but not limited to \lgc)~\cite{ippolito2019automatic, mitchell2023detectgpt, su-etal-2023-detectllm}.
For example, \citet{mitchell2023detectgpt} proposed DetectGPT to detect LLM-generated text. Technically, DetectGPT utilizes the log probabilities to calculate the perplexity. The authors evaluated DetectGPT using the AUC metric on 500 LLM-generated samples and 500 human-written samples. In addition, they evaluated the generalization capability of DetectGPT by alternately using GPT-J-6B, GPT-Neo-2.7B, and GPT-2-1.5B as the code generation model and the perplexity calculation model.
\citet{su-etal-2023-detectllm} proposed two methods to calculate the perplexity. The first is \underline{L}og-Likelihood \underline{L}og-Rank \underline{R}atio (LRR) and the second is \underline{N}ormalized \underline{P}erturbed Log-\underline{R}ank (NPR). The authors evaluated LRR and NPR using the AUC metric on 300 LLM-generated and 300 human-written samples. They also compared LRR and NPR against baselines in terms of detection speed.  

With the increasing popularity of \lgc, \per has been commonly used in detecting \lgc from \hac~\cite{shi2024between, xu2024detecting}. 
\citet{xu2024detecting} proposed a method using a variant of probability to calculate the perplexity. They evaluated their methods using the AUC, FPR, and TPR metrics on 5,214 LLM-generated and 5,214 human-authored code snippets. They also evaluated the influence of models used to calculate perplexity.
\citet{shi2024between} proposed a method namely DetectCodeGPT to detect \lgc from \hac. DetectCodeGPT calculates the perplexity based on NPR. They evaluated DetectCodeGPT using the AUC metric on 500 LLM-generated and 500 human-authored code snippets. They also evaluated the generalization capability of DetectCodeGPT by alternately using Incoder-1B, Phi-1B, StarCoder-3B, WiardCoder-3B, CodeGen2-3.7B, and GodeLlama-7B as the code generation model and the perplexity calculation model.
These studies focus on proposing new perplexity-based detection methods and evaluating overall detection accuracy. As a result, they only partially consider other detection methods and pay limited attention to broader evaluation aspects, leading to limitations in both evaluation objectives and dimensions. To address this gap, this study conducts a systematic evaluation of feature-based, pre-training-based, and perplexity-based methods from multiple perspectives, including detection accuracy (language, difficulty, scale), detection speed, and generalization capability.

Overall, \per is a popular method for detecting \lgc from \hac but its evaluation is limited. In this article, we conduct a large-scale investigation on the efficacy of \per from multiple aspects.

\section{Motivating Examples}
\label{SEC:MOT}
This section describes three examples to motivate this investigation.

\subsection{Code Authenticity}
Although \per has been demonstrated useful in detecting \lgc,  the \lgc used in previous work is generated by ``tiny'' \llm, such as CodeGen2-3.7B~\cite{shi2024between}. Their effectiveness for larger code generators is unknown. 

\begin{figure}[htbp]
    \begin{center}
    \includegraphics[width=1\textwidth]{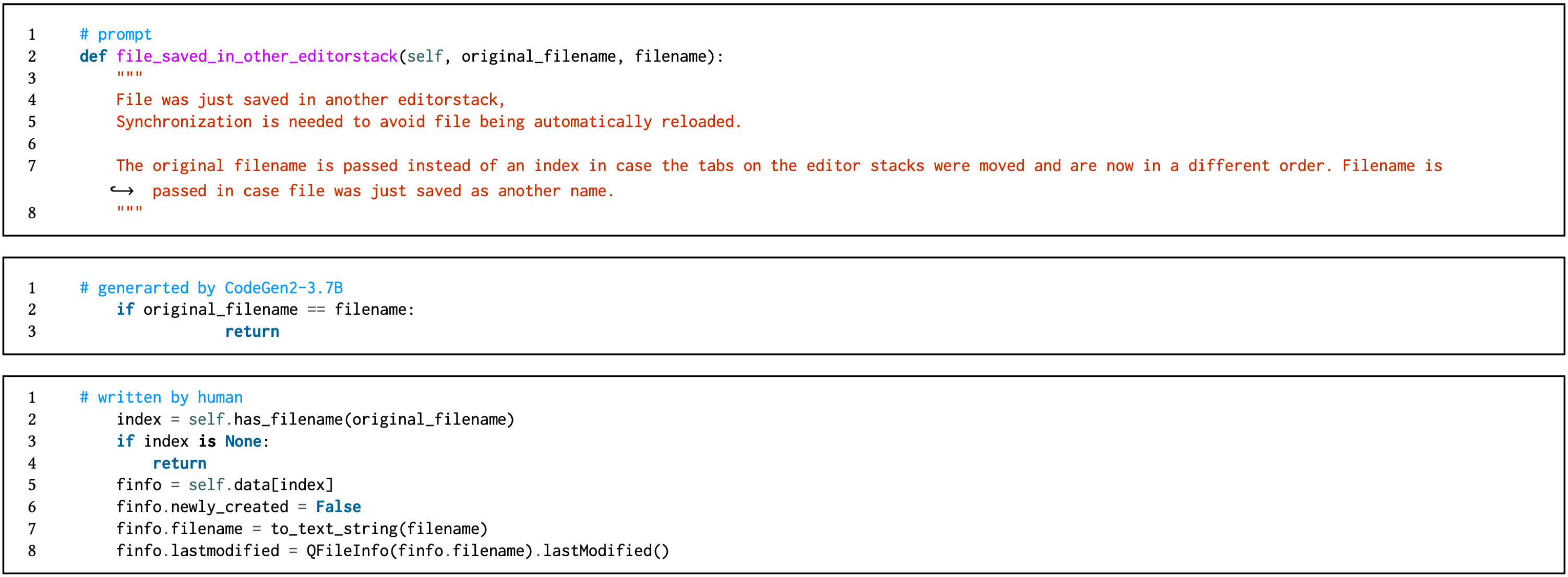}
    \end{center}
    \caption{An example of the \lgc (CodeGen2-3.7B) and \hac}
\label{FIG:example-1}
\vspace{1.0ex}
\end{figure}

\Cref{FIG:example-1} presents examples of \lgc and \hac. The prompt indicates that the function to be generated is to synchronize the file state across editor stacks to prevent unnecessary reloads after a save. 
The prompt is the input of CodeGen2-3.7B (shown in the first block).
The \lgc is the solution generated by CodeGen2-3.7B (shown in the second block), and the \hac is the solution provided by a human developer (shown in the third block). 
The \lgc is an incorrect implementation of this functionality and differs significantly from \hac. 
In fact, the \lgc only returns when the file was not saved under a new name, which is inconsistent with the intended functionality.
While \hac avoids unnecessary reloads by updating the file state: it locates the original filename and synchronizes its metadata, including the creation status, filename, and newest modified time. 
Although it is easy to detect such \lgc from \hac, few developers use such \lgc for their code solutions, as it is far from addressing real-world problems.

When \per uses the same model for code generation and detection, it achieves low perplexities of \lgc. 
This is because the model would assign a high prediction probability to each token when it perceives the code generated by itself. 
However, in real-world scenarios, the detection model does not have clues to know which model is used to generate code early.
Moreover, these \llm used for generating code have hundreds of billions of parameters, making them challenging to deploy locally to calculate perplexities.

\subsection{Language Diversity}
\begin{figure}[htbp]
    \begin{center}
    \includegraphics[width=0.99\textwidth]{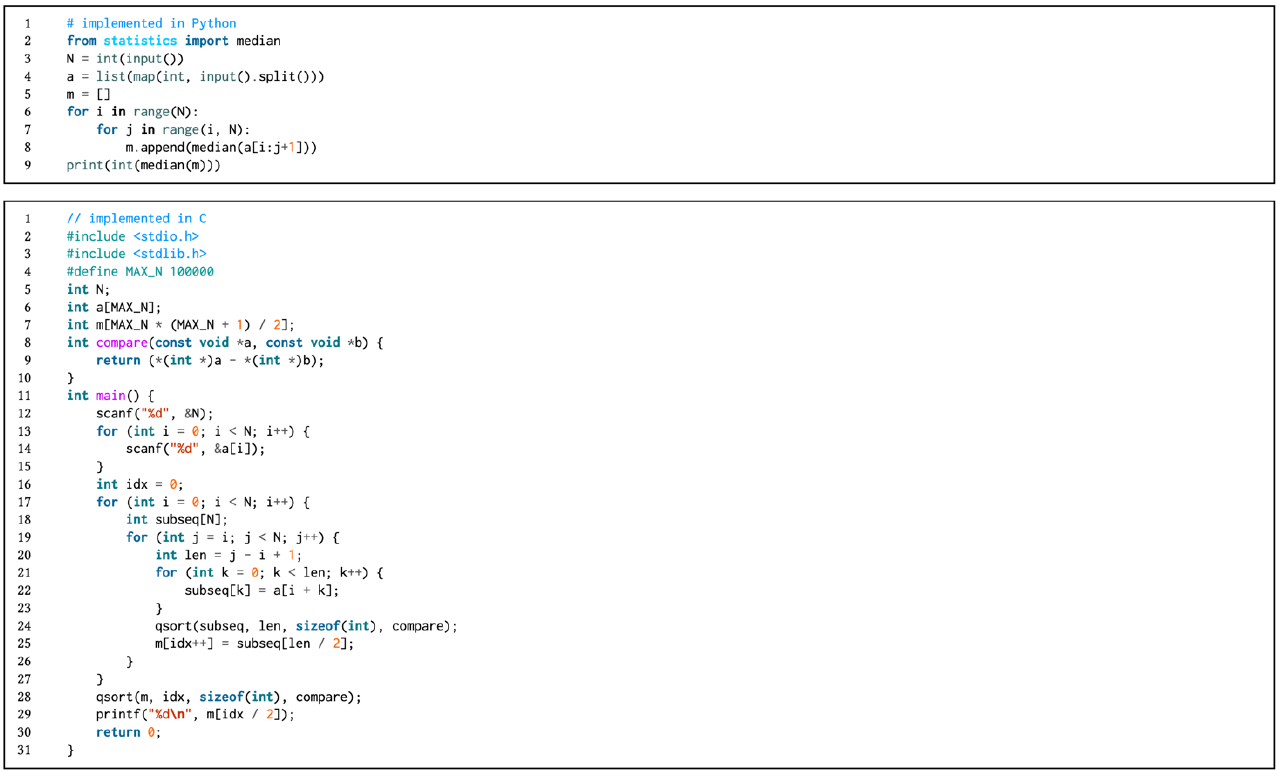}
    \end{center}
    \caption{An example of code achieving the same task in Python and C}
\label{FIG:example-2}
\end{figure}

\per has been evaluated in limited programming languages so far.
However, the code stylometry of different programming languages varies greatly. 
\Cref{FIG:example-2} presents two pieces of code generated by GPT-3.5, aiming to calculate the median of all subarrays of a given array and then calculate the median of these medians.

As illustrated in \Cref{FIG:example-2}, the code stylometries differ in Python and C, particularly in their logic and scale. 
Python has rich built-in libraries. 
For example, ``\texttt{statistics.comedian}'' is a tool library in Python. It can directly calculate the median without the need for manual sorting and selection. 
While C does not have a direct median calculation function, therefore, we use ``\texttt{qsort}'' to sort and manually calculate the median. 
In fact, many high-level functions must be manually implemented in C, such as inserting or deleting elements in arrays. However, such functions can be automatically handled by the built-in methods of lists in Python. 
In addition, Python is a dynamically typed programming language that does not require declaring variable types, making the code more concise. While C is a statically typed programming language where each variable type needs to be declared. 
Moreover, manipulating elements such as list slicing is easy in Python. For example, ``\texttt{a[i: j+1]}'' returns subarrays without manual calculation and traversal. While C requires manually managing memory and arrays, which means that developers need to handle each element carefully.

Due to differences such as the built-in library support and design principles, code written in different programming languages is distinguishable. In addition to programming languages, the code used to solve puzzles of different difficulty and the code of different scales are distinguishable in terms of logic and complexity.

\subsection{Generalization Capability}
\begin{figure}[htbp]
    \begin{center}
    \includegraphics[width=0.99\textwidth]{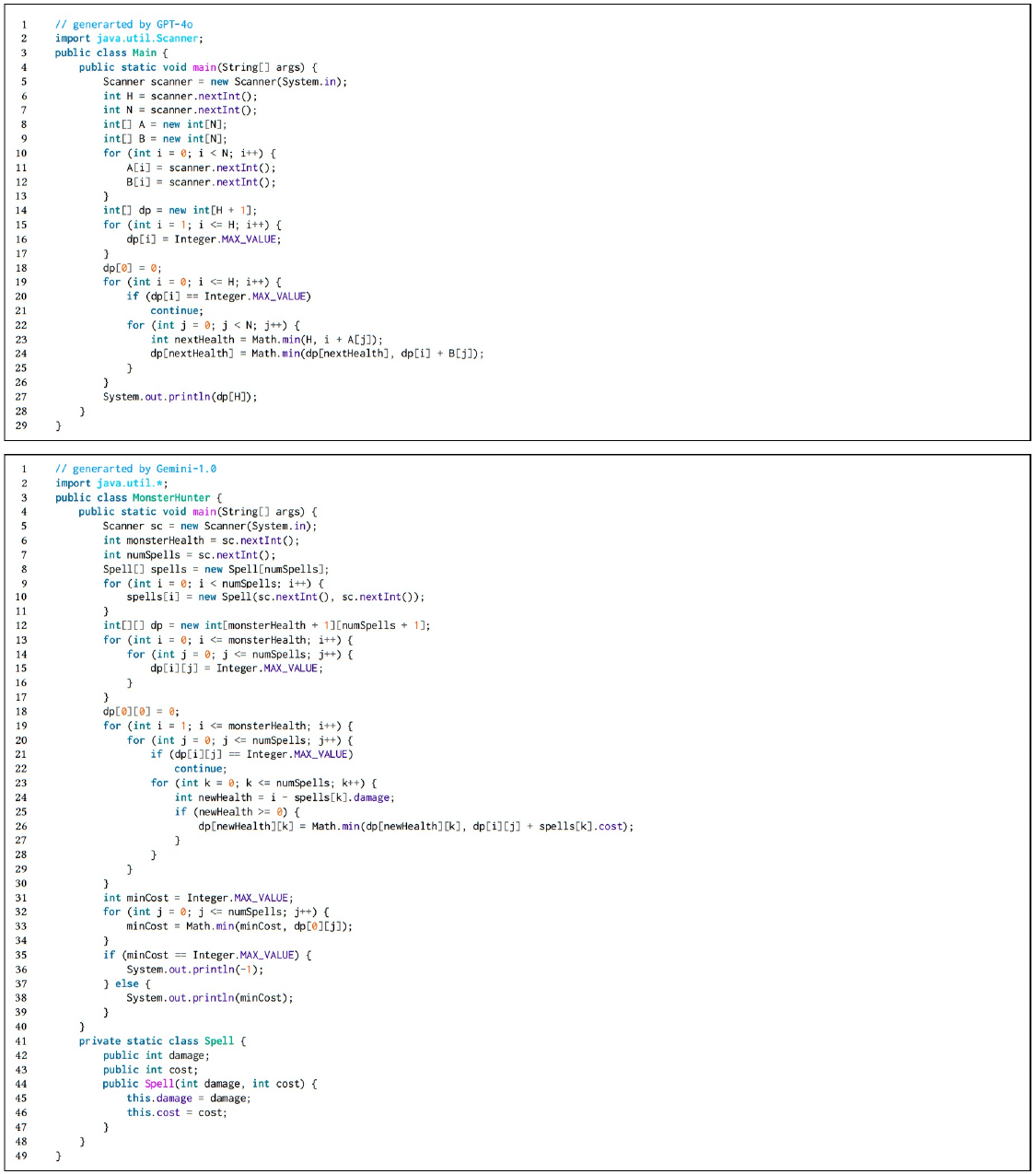}
    \end{center}
    \caption{An example of code achieving the same task generated by GPT-4o and Gemini-1.0}
\label{FIG:example-3}
\end{figure}

There are many \llm available for generating code, such as GPT, Copilot, and Gemini. 
Due to differences in training data, model architecture, optimization objectives, generation strategies, etc., the code generated by these \llm is different. 
\Cref{FIG:example-3} presents two Java code snippets generated by GPT-4o and Gemini-1.0 solving the same programming puzzle. 
The puzzle is a dynamic programming problem that is similar to the unbounded knapsack problem. 
This problem can be described as follows. Given the monster's initial health $H$ and $N$ different spells, each spell has a fixed damage $A_{i}$ and a magic point cost $B_{i}$. 
The objective is to reduce the monster's health to 0 or below by casting spells while minimizing the total magic points consumed.

As shown in \Cref{FIG:example-3}, although both pieces of code aim to reduce the monster's health to 0 with minimal loss, the two \llm have different implementations. 
GPT-4o starts to generate code from the point of view of damage, with the aim of achieving total damage $H$ at the minimum cost. 
Gemini-1.0 starts to generate code from the perspective of the monster's health, aiming to reduce the monster's health to 0 with the minimum cost. 
The former is to intuitively accumulate damage in the forward direction, and the latter is to reduce the monster's health in the reverse direction. 

In addition to the differences mentioned above, the abstraction levels are different. 
The solution generated by Gemini-1.0 is detailed and complex. Gemini-1.0 uses a two-dimensional array `\texttt{dp}' to simultaneously record health values and spell combinations. 
GPT-4o uses only a one-bit array `\texttt{dp}' to record the minimum cost of achieving each damage value. 
When updating the state, Gemini-1.0 considers all spell combinations in each update, making it more complex, while GPT-4o directly updates the state by traversing damage values and spell methods.
In addition, Gemini-1.0 considers boundary conditions and special cases, such as returning an outlier of ``\texttt{-1}'' when the target cannot be reached.
Unfortunately, GPT-4o fails to handle these boundary conditions, reducing its robustness. 

In short, GPT-4o focuses on core logic, while Gemini-1.0 focuses on practical usages, boundary conditions, and robustness regarding this example. 
This example motivates us to investigate the generalization capability of \per when the code is generated by different \llm. In addition to generalization, the detection speed of \per is important, as it influences user experience.

\section{Research Methodology}
\label{SEC:MED}
This section elaborates on the experimental designs to investigate the efficacy of \per in detecting \lgc.

\subsection{Research Questions}
Existing work focuses on evaluating the detection accuracy of \per, however, there are many aspects to consider when using it in practice. In this article, we investigate the efficacy of \per in detecting \lgc in a wider range of realistic evaluation settings.
\smallskip

\begin{itemize}
    \item [$\text{\textbf{RQ}}_\textbf{1}$:] \textbf{Is \per an effective method in detecting \lgc?: \textbf{Language} ($\text{\textbf{RQ}}_{\textbf{1.1}}$), \textbf{Difficulty} ($\text{\textbf{RQ}}_{\textbf{1.2}}$), and \textbf{Scale} ($\text{\textbf{RQ}}_{\textbf{1.3}}$)}\\     
    \emph{Motivation:}
    Detection accuracy is crucial in detecting \lgc, while it might not be consistently good between situations, \eg an effective method for C++ might not be suitable for Python.
    We propose $\text{RQ}_1$ to compare \per against state-of-the-art methods regarding detection accuracy from three sub-aspects: 
    \begin{inparaenum}[(1)]
        \item \emph{programming language} of the implementation: C, C\#, C++, Go, Java, Python, Ruby, and Rust;
        \item \emph{degree of difficulty} of the puzzle: easy, medium, and hard; and
        \item \emph{scale of solution} to the puzzle: small, middle, and large.
    \end{inparaenum}
\end{itemize}
\smallskip

\begin{itemize}
    \item [$\text{\textbf{RQ}}_\textbf{2}$:] \textbf{Is \per an efficient method in detecting \lgc?}\\ 
    \emph{Motivation:} 
    Detection speed affects user experience in just-in-time scenarios such as time-limited programming contests, and is important for large-scale applications such as batch scanning of software.
    According to the detection mechanism, \per applies perturbation strategies to generate dozens of perturbation samples and uses an LLM to calculate the probability or rank of each token for both the original sample and perturbation samples~\cite{shi2024between, su-etal-2023-detectllm}. 
    Such a two-phase detection process might be time-consuming. 
    In $\text{RQ}_2$, we compare \per with feature- and pre-training-based methods regarding detection speed.
\end{itemize}
\smallskip

\begin{itemize}
    \item [$\text{\textbf{RQ}}_\textbf{3}$:] \textbf{Does \per have the generalization capability?}\\ 
    \emph{Motivation:} 
    Developers utilize various \llm to assist in programming. 
    Due to differences in training data, model architecture, and generation strategies, the code generated by different \llm varies significantly. 
    An optimal detection method should be capable of detecting the code generated by different \llm. 
    In $\text{RQ}_3$, we investigate the generalization capability of \per in detecting \lgc.
\end{itemize}
\smallskip

\subsection{Data Preparation}
Although there are datasets~\cite{shi2024between, bukhari2023distinguishing, idialu2024whodunit} that work to detect \lgc from \hac, none of them is suitable to address all research questions. 
These datasets have the following limitations: (1) they support only Python and C (limited diversity in languages); (2) their \lgc code is generated from a single LLM or a single family of LLMs (limited capability in generalization); and (3) only one dataset~\cite{idialu2024whodunit} provides difficulty scores of puzzles, (limited degrees in difficulty).
Therefore, we developed a new dataset based on CodeNet~\cite{puri2021codenet}. CodeNet was developed by IBM to promote AI technologies in the code domain and has been widely used for many studies~\cite{xu2024detecting, agarwal2021using, goel2022cross, pan2024lost, naik2022probing, dinh2024large}. CodeNet comprises a large number of submissions from real-world developers. More than half of these submissions are functionally correct (accepted), and more than 90\% are successfully compiled and executed (executable). The high quality of CodeNet ensures our \hac serves as a reliable baseline.

\begin{figure}[htbp]
    \begin{center}
    \includegraphics[width=0.99\textwidth]{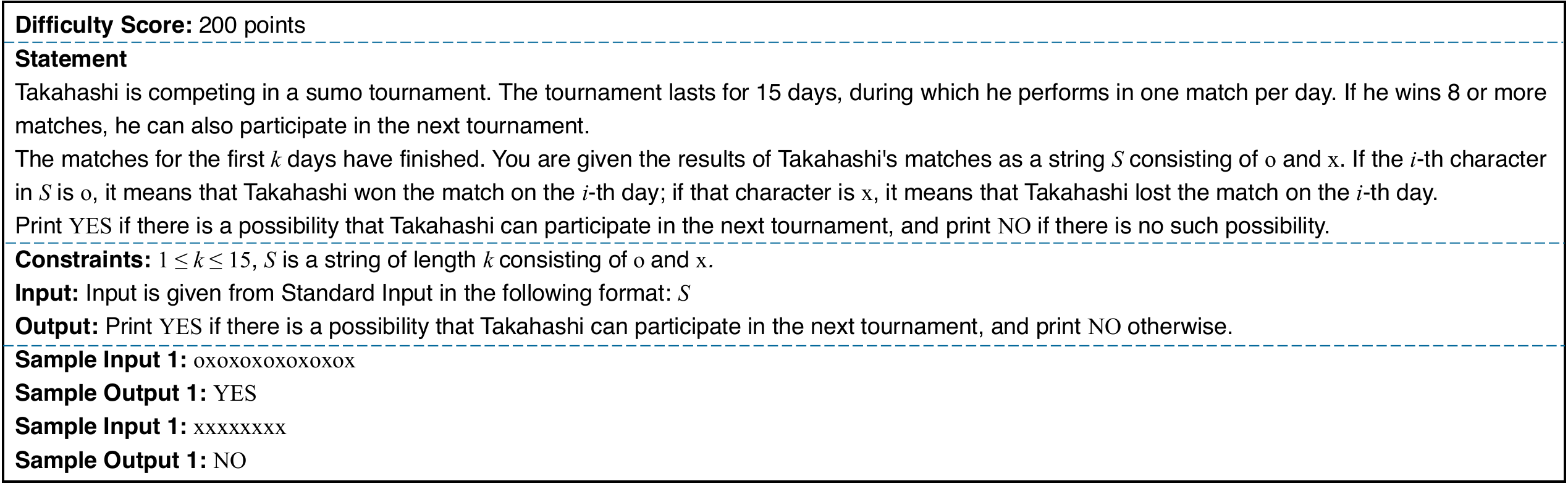}
    \end{center}
    \caption{An example of a programming puzzle in CodeNet}
\label{FIG:dataset}
\end{figure}

As shown in \Cref{FIG:dataset}, a programming puzzle provided by CodeNet has seven basic elements: (1) \emph{difficulty score}, (2) \emph{statement}, (3) \emph{constraint}, (4) \emph{input}, (5) \emph{output}, (6) \emph{sample input}, and (7) \emph{sample output}. The difficulty score measures the difficulty level of the puzzle. The statement describes what the puzzle is. The constraint describes the specific constraints on the variables. The input and output describe the requirements for input and output, respectively. Finally, the sample input and sample output provide the input and output of a specific example, respectively. Based on CodeNet, we developed our dataset through the following five steps.
\smallskip

\begin{figure}[htbp]
    \begin{center}
    \includegraphics[width=0.99\textwidth]{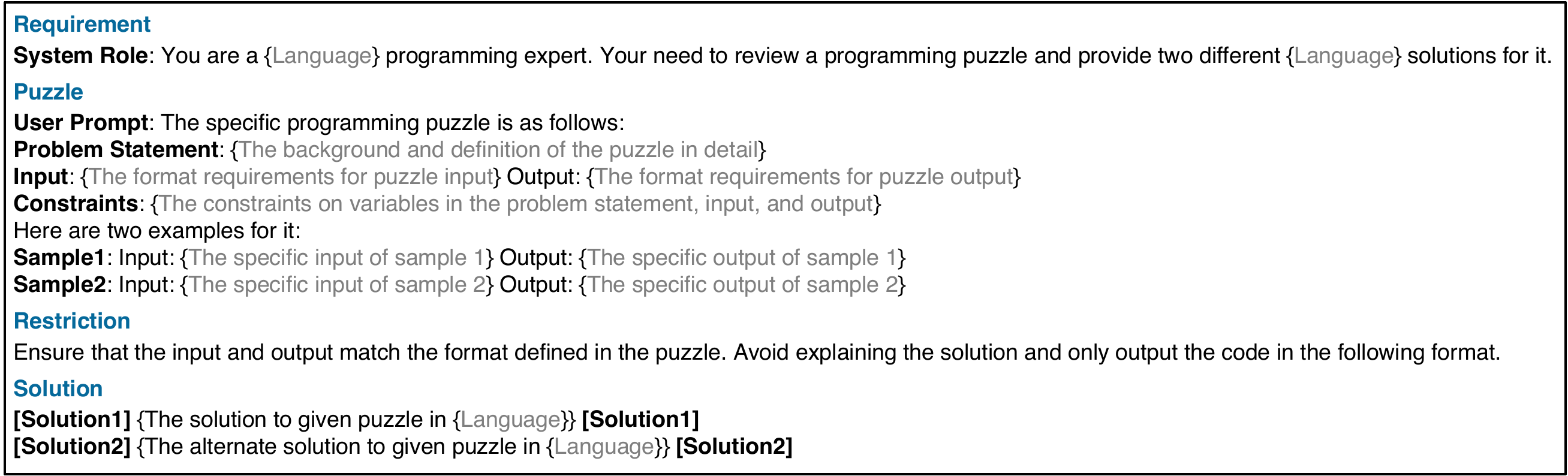}
    \end{center}
    \caption{The prompts used for generating \lgc solutions for programming puzzles}
\label{FIG:prompt}
\end{figure}

\begin{itemize}
    \item \textbf{Step-1: Problem selection.}
    To build an appropriate dataset for evaluation, we excluded the programming problem whose (1) statement is not in English as it is the most popular language; (2) statement contains images or tables as they can hardly be extracted and input into \llm; (3) solutions are all less than 30 LOC (Lines Of Code) as they are mainly for trivial puzzles; (4) solutions are incomplete regarding all the eight programming languages as they constitute the \hac to be evaluated in $\text{RQ}_{1.1}$; and (5) structure is not complete, \ie \emph{statement}, \emph{input}, \emph{output}, \emph{constraints}, \emph{difficulty}, and \emph{samples} ($\geq 2$). After employing these criteria, our dataset includes 729 programming puzzles.
    \smallskip

    \item \textbf{Step-2: \hac collection.}
    For each puzzle selected in \emph{Step-1}, we randomly took two code solutions for each of the eight programming languages from CodeNet. The code solutions were submitted by human developers, including correct, incorrect, and rejected solutions. Finally, we had 16 \hac solutions for each puzzle.
    \smallskip

    \item \textbf{Step-3: \lgc generation.}
    We generated \lgc for programming puzzles using OpenAI GPT-3.5 (gpt-3.5-turbo-0125). The prompts for code generation are shown in \Cref{FIG:prompt}. There are two solutions for each puzzle regarding each of the eight programming languages. Moreover, to address $\text{RQ}_3$, we used the prompts from \Cref{FIG:prompt} and input them into GPT-4o, Gemini-1.0, and Llama-3.1, with each model generating 500 solutions. Note that the following two steps are performed on the code solutions that have been generated by GPT-3.5.
    \smallskip

    \item \textbf{Step-4: Classification according to the difficulty of puzzles.}
    We categorized the puzzles based on their difficulty degrees and set two thresholds for the difficulty score to ensure that each group has the same number of puzzles. The detailed settings are: \emph{easy} for $\leq 200$, \emph{medium} for $> 200$ and $\leq 500$, and \emph{hard} for $> 500$.
    \smallskip

    \item \textbf{Step-5: Classification according to the scale of solutions.} 
    We categorized the code solutions based on their scale (LOC) and set two thresholds to ensure that each group has the same number of solutions. The detailed settings are: \emph{small} for $\leq 20$, \emph{medium} for $> 20$ and $\leq 50$, and \emph{large} for $> 50$.
\end{itemize}

\smallskip

Finally, we obtained a dataset containing function-level \lgc and \hac that solve the same programming puzzles, which is used for the experimental evaluations. We perform the evaluation on function-level code because (1) due to the current limitations of \lgc in handling complex programming tasks, developers usually use it to assist in implementing functions rather than entire software repositories. As a result, repo-level detection also needs to be performed at the function level, and (2) to the best of our knowledge, existing \lgc detection methods are all designed for function-level detection. Therefore, we evaluate them using the function-level code to ensure a fair evaluation.

\subsection{Studied Methods}
We compared \per against feature- and pre-training-based methods. 

\subsubsection{Feature-based Methods}
Feature-based methods require professionals to devise appropriate features of code, \eg lexical and syntactic features, and input these features into machine learning classifiers for detection. We investigate two feature-based methods as follows.
\smallskip

\begin{itemize}
    \item \textbf{Idialu} et al.~\cite{idialu2024whodunit} takes a combination of layout, syntactic, and lexical features of Python programs as the input, where layout features refer to code formatting such as indentation and spacing, lexical features are derived from code tokens such as keywords and literals, and syntactic features capture the structural patterns and relationships between code elements. Idialu uses XGBoost to construct the detection method.\smallskip
    
    \item \textbf{Bukhari} et al.~\cite{bukhari2023distinguishing} takes a combination of lexical and syntactic features of C programs as the input. The lexical features are derived from the feature set introduced by Caliskan-Islam~\cite{caliskan2015anonymizing} for code stylometry. The syntactic features are extracted from abstract syntax trees. Bukhari uses Random Forest, SVM, KNN, and XGBoost to build the detection methods. XGBoost outperforms the other classifiers in terms of detection performance.
\end{itemize}

\subsubsection{Pre-Training-based Method}
Pre-training-based methods take code snippets as input, convert them into tokens, embed them as vectors, and leverage a pre-trained model for classification. We investigate one pre-training-based method as follows.
\smallskip

\begin{itemize}
    \item \textbf{GPTSniffer}~\cite{nguyen2024gptsniffer} designs rewriting rules (\eg rewriting class names) to create diverse derivations of the original code samples. These derivations are then used to fine-tune a pre-trained model, \ie CodeBERT, as the detector.
    To perform detection, GPTSniffer transforms the input code into a sequence of independent tokens, which are then fed into the fine-tuned CodeBERT to predict whether the code is \lgc.  
\end{itemize}

\subsubsection{\per}
Eight popular perplexity-based methods are as follows. The first five calculate the perplexity based on the original sample without perturbations. The last three require applying other \llm to perturb the original sample and calculate the perplexity based on the discrepancies before and after the perturbation.
\smallskip

\begin{itemize}
    \item \textbf{Log-p(x)}~\cite{solaiman2019release} utilizes the average token-wise log probability to calculate the perplexity. The log probability is the logarithm of the probability of the next token (ground truth) based on previous tokens, and the probability is calculated by a language model. This method assumes that \lgc has a larger log-p(x) than \hac. The calculation of log-p(x) is shown in \Cref{EQ:Log-p(x)}, where $ p_\theta(x_i \mid x_{<i}) $ is the probability of token $ x_i $ conditioned on the previous tokens, $t$ is the number of tokens in the sequence. 
    \smallskip
    \begin{equation}
    \label{EQ:Log-p(x)}
         \textrm{Log-p(x)} = \frac{1}{t} \sum_{i=1}^{t} \log p_\theta(x_i \mid x_{<i})
    \end{equation}
    \smallskip

    \item \textbf{Entropy}~\cite{gehrmann2019gltr} uses the average entropy to calculate the perplexity. Entropy is the sum of the probability multiplied by the negative logarithm of the probability of each token. This method assumes that \lgc has a larger entropy than \hac. The calculation of entropy is shown in \Cref{EQ:Entropy}, where $ p_\theta(x_i \mid x_{<i}) $ is the probability of token $ x_i $ conditioned on the previous tokens, $t$ is the number of tokens in the sequence. 
    \smallskip
    \begin{equation}
    \label{EQ:Entropy}
         \textrm{Entropy} = -\frac{1}{t} \sum_{i=1}^{t} p_\theta(x_i \mid x_{<i}) \log p_\theta(x_i \mid x_{<i})
    \end{equation}
    \smallskip

    \item \textbf{Log-Rank/Rank}~\cite{gehrmann2019gltr, ippolito2019automatic} utilizes the average log-rank/rank of each token to calculate the perplexity. Log-Rank/Rank is the logarithm of rank/rank of the next token in the vocabulary list. The vocabulary list is sorted according to the probability of each token given the previous tokens, and the probability is calculated by a language model. This method assumes that \lgc has a lower log-rank/rank than \hac. The calculations of Log-Rank and Rank are shown in \Cref{EQ:Log-Rank} and \Cref{EQ:Rank} respectively, where $ r_\theta(x_i \mid x_{<i}) $ is the rank of token $ x_i $ conditioned on the previous tokens, $t$ is the number of tokens in the sequence.   
        \smallskip
    \begin{equation}
    \label{EQ:Log-Rank}
         \textrm{Log-Rank} = \frac{1}{t} \sum_{i=1}^{t} \log r_\theta(x_i \mid x_{<i})
    \end{equation}

    \begin{equation}
    \label{EQ:Rank}
         \textrm{Rank} = \frac{1}{t} \sum_{i=1}^{t} r_\theta(x_i \mid x_{<i})
    \end{equation}
    \smallskip

    \item \textbf{LRR}~\cite{su-etal-2023-detectllm} divides log-likelihood by log-rank to calculate the perplexity. Log-likelihood is the average token-wise log probability, representing the absolute confidence for the true token. Log-rank is the average token-wise log rank, representing the relative confidence for the true token. For \lgc, log-rank is more discernible than the log-likelihood. This method assumes that \lgc has a larger LRR than \hac. The calculation of LRR is shown in \Cref{EQ:LRR}, where $ p_\theta(x_i \mid x_{<i}) $ is the probability of token $ x_i $ conditioned on the previous tokens, $ r_\theta(x_i \mid x_{<i}) $ is the rank of token $ x_i $ conditioned on the previous tokens, $t$ is the number of tokens in the sequence. 
    
    \smallskip
    \begin{equation}
    \label{EQ:LRR}
         \textrm{LRR} = - \frac{\sum_{i=1}^{t} \log p_\theta(x_i \mid x_{<i})}{\sum_{i=1}^{t} \log r_\theta(x_i \mid x_{<i})}
    \end{equation}
    \smallskip
    
    \item \textbf{DetectGPT}~\cite{mitchell2023detectgpt} leverages the discrepancy between the log probability of the original sample (a collection of tokens) and the average log probability of the perturbed samples to calculate the perplexity, normalized by the standard deviation of the log probability of the perturbed samples. The perturbed samples are obtained by masking and completing some tokens. This method assumes that \lgc has a greater perturbation discrepancy as it is more sensitive to perturbations than \hac. The calculation of DetectGPT is shown in \Cref{EQ:DetectGPT}, where $x$ is the original sample, $\tilde{x}$ is the perturbed sample, $ \log p_\theta(x) $ is the log probability of the original sample $x$, $\log p_\theta(\tilde{x}_{i}) $ is the log probability of the perturbed sample $\tilde{x}_{i}$, $k$ is the number of perturbed samples. 
    \smallskip
    \begin{equation}
    \label{EQ:DetectGPT}
         \textrm{DetectGPT} = \frac{ \log p_\theta(x)-\frac{1}{k}\sum_{i=1}^{k}\log p_\theta(\tilde{x_{i}})}{\sqrt{\frac{1}{k-1}\sum_{p=1}^{k} {(\log p_\theta(\tilde{x}_{p})-\frac{1}{k}\sum_{i=1}^{k}\log p_\theta(\tilde{x}_{i}))}^{2} }}
    \end{equation}
    \smallskip

    \item \textbf{NPR}~\cite{su-etal-2023-detectllm} divides the average log-rank of the perturbed samples by the log-rank of the original sample to calculate the perplexity. The perturbed samples are obtained by replacing some tokens. This method assumes that \lgc has a larger NPR as it is more susceptible to perturbations than \hac. The calculation of NPR is shown in \Cref{EQ:NPR}, where $x$ is the original sample, $\tilde{x}$ is the perturbed sample, $ \log r_\theta(x) $ is the log rank of the original sample $x$, $\log r_\theta(\tilde{x}_{p}) $ is the log rank of the perturbed sample $\tilde{x}_{p}$, $k$ is the number of perturbed samples.  
    \smallskip
    \begin{equation}
    \label{EQ:NPR}
         \textrm{NPR} =  \frac{ \frac{1}{k}\sum_{p=1}^{k} \log r_\theta( \tilde{x}_{p})}{\log r_\theta(x)}
    \end{equation}
    \smallskip

    \item \textbf{DetectCodeGPT}~\cite{shi2024between} utilizes the discrepancy between the NPR of the original sample and the average NPR of the perturbed samples to calculate the perplexity. The perturbed samples are obtained by adding spaces and new lines to the original sample. This method assumes that the average NPR of perturbed \lgc drops more than \hac. The calculation of NPR is shown in \Cref{EQ:DetectCodeGPT}, where $x$ is the original sample, $\tilde{x}$ is the perturbed sample, $ \log r_\theta(x) $ is the log rank of the original sample $x$, $\log r_\theta(\tilde{x}_{i}) $ is the log rank of the perturbed sample $\tilde{x}_{i}$, $k$ is the number of perturbed samples.   
    \smallskip
    \begin{equation}
    \label{EQ:DetectCodeGPT}
         \textrm{DetectCodeGPT} =  \frac{ \frac{1}{k}\sum_{p=1}^{k} \log r_\theta( \tilde{x}_{p})}{\log r_\theta(x)} - \frac{1}{k} \sum_{i=1}^{k} \frac{ \frac{1}{k} \sum_{p=1}^{k} \log r_\theta( \tilde{x}_{p})}{\log r_\theta(\tilde{x}_{i})}
    \end{equation}
    \smallskip
    
\end{itemize}

When selecting experimental subjects, we included several detection methods that were not originally proposed to detect \lgc, \eg DetectGPT and NPR. They were proposed to detect LLM-generated text. Since source code is one type of text, these detection methods are also suitable for detecting \lgc. This setting was also used in previous related work~\cite{shi2024between, suh2025empirical}. Although \per is an unsupervised method, we compared it against supervised methods because \citet{shi2024between} reported that \per outperformed supervised methods. We excluded methods whose implementation details are unclear. For example, we excluded AIgCode Detector~\cite{xu2024detecting} because it did not specify how perturbations are performed.

\subsection{Experimental Settings}
Following related work~\cite{shi2024between, idialu2024whodunit, mitchell2023detectgpt, su-etal-2023-detectllm}, we use the AUC (Area Under the Curve) to evaluate each detection method. AUC stands for the area under ROC (Receiver Operating Characteristic curve). Given a set of TPR (True Positive Rates) and FPR (False Positive Rates) across different thresholds, the calculation of AUC is as follows.

\smallskip
\begin{equation}
\label{EQ:e1}
     \textrm{AUC} = \int_{0}^{1}\textrm{TPR}(t)d\textrm{FPR}(t)
\end{equation}
\smallskip

where $t$ denotes the varying threshold values.

For $\text{RQ}_{1.1}$, we evaluated \per and pre-training-based methods in all eight programming languages. We evaluated feature-based methods on specific languages because the features are designed for those languages only.
Due to the time and cost limit, we tested 500 pairs of human-LLM code snippets for each aspect regarding each research question. For $\text{RQ}_{1}$, we tested 500 pairs of human-LLM code snippets for each programming language, difficulty level, and scale. For $\text{RQ}_{2}$ and $\text{RQ}_{3}$, we tested 500 pairs of human-LLM code snippets for each LLM. We randomly selected 500 pairs of human-LLM code snippets from the dataset each time. We selected 500 pairs for the reason of budget and time costs, and such a scale is comparable to the prior related work~\citet{shi2024between} (500) and \citet{suh2025empirical} (400).
Following \citet{shi2024between}, we utilized CodeT5-770M to generate perturbed code snippets for DetectGPT. We added spaces and empty lines to generate perturbed code snippets for NPR and DetectCodeGPT. We generated 50 perturbed samples for each code sample. 
Code-specific LLMs are pre-trained using code corpora, which equips them with a deeper understanding of programming languages. We use code-specific LLMs to compute the perplexity of code snippets, as they are more capable of understanding code than general LLMs.
Following \citet{shi2024between}, we utilized Incoder (1.3B), Phi-1 (1.3B), StarCoder (3B), WizardCoder (3B), CodeGen2 (3.7B), and CodeLlama (7B) to calculate the perplexity for each \per.
Table~\ref{tab:LLMRole} summarizes the specific usage of all LLMs involved in our study.
We carried out the experiments using 1 NVIDIA A100-PCIE-40GB.

\begin{table}[ht]
\centering
\footnotesize
\caption{Summary of LLM usage}
\begin{tabular}{l|l}
\hline
\textbf{LLM} & \textbf{Usage} \\
\hline
GPT3.5 & To generate 11,664 \lgc solutions for all the evaluation aspects. \\
\hline
GPT-4o & \multirow{3}{*}{To generate 500 \lgc solutions for the generalization capability.}\\ 
Gemini-1.0 & \\
Llama-3.1 & \\
\hline
Incoder (1.3B) & \multirow{6}{*}{To calculate perplexity values for all methods in \per.}\\
Phi-1  (1.3B)& \\
StarCoder (3B)& \\
CodeGen2 (3B)& \\
WizCoder (3.7B)& \\
CodeLlama (7B)& \\
\hline
CodeT5 (770M) & To generate perturbed code snippets for DetectGPT.\\
\hline
\end{tabular}

\label{tab:LLMRole}
\end{table}

\section{Results and Analysis}
\label{SEC:RES}
This section reports on results and analysis to answer research questions.

\subsection{$\text{RQ}_\text{1.1}$: Programming Language}
\Cref{tab:performance-perplexity} and \Cref{tab:performance-feature} show the AUC of the three groups of detection methods from the perspective of programming languages. The code LLMs listed in \Cref{tab:performance-perplexity} are used to calculate perplexity values. Since feature-based detection methods are designed for specific languages, we provide results only in Python (Idialu) and C (Bukhari) in \Cref{tab:performance-feature}. The notable results are highlighted in blue (AUC:\colorbox{gray1}{$90\%-100\%$};\colorbox{gray2}{$80\%-90\%$};\colorbox{gray3}{$70\%-80\%$}).

{\footnotesize
\begin{longtable}[ht]{llccccc|ccc}
\caption{The AUC of \per in terms of programming languages}\\
\label{tab:performance-perplexity}\\ 

    \hline
    \multirow{2}{*}{\textbf{Language}} & \multirow{2}{*}{\textbf{Code LLM}} & \multicolumn{5}{c}{\textbf{without perturbation}} & \multicolumn{3}{c}{\textbf{with perturbation}} \\
    \cmidrule{3-10}
    & & log-p & Ent & Rank & L-Rank & LRR & DGPT & NPR & DCGPT \\
    \hline
    \endfirsthead

    \multicolumn{10}{c}%
    {{\bfseries Table \thetable\ continued from previous page}} \\
    \hline
    \multirow{2}{*}{\textbf{Language}} & \multirow{2}{*}{\textbf{Code LLM}} & \multicolumn{5}{c}{\textbf{without perturbation}} & \multicolumn{3}{c}{\textbf{with perturbation}} \\
    \cmidrule{3-10}
    & & log-p & Ent & Rank & L-Rank & LRR & DGPT & NPR & DCGPT \\
    \hline
    \endhead

    \multirow{6}{*}{\shortstack{C}} 
    & Incoder & 56.43 & 38.55 & 19.07 & 59.98 & 64.80 & 57.31 & \cellcolor{gray2}84.98 & \cellcolor{gray2}84.80\\
    & Phi-1 & \cellcolor{gray2}83.66 & 36.19 & 59.41 & \cellcolor{gray3}77.13 & 25.84 & 61.66 & \cellcolor{gray1}90.17 & \cellcolor{gray1}\textbf{90.40}\\
    & StarCoder & 65.43 & 48.50 & 49.85 & 66.72 & 58.72 & 56.32 & \cellcolor{gray2}86.47 & \cellcolor{gray2}85.70\\
    & CodeGen2 & \cellcolor{gray3}74.30 & 39.72 & 52.23 & 72.12 & 53.33 & 56.40 & \cellcolor{gray2}88.51 & \cellcolor{gray2}87.08\\
    & WizCoder & \cellcolor{gray2}80.87 & 30.02 & 58.61 & \cellcolor{gray3}77.82 & 53.16 & 58.65 & \cellcolor{gray2}88.73 & \cellcolor{gray2}89.09\\
    & CodeLlama & \cellcolor{gray3}74.75 & 41.99 & 57.49 & \cellcolor{gray3}70.25 & 58.83 & 57.27 & \cellcolor{gray2}85.05 & \cellcolor{gray2}84.95\\
    \cmidrule{2-10}
    & Average & \cellcolor{gray3}72.57 & 39.16 & 49.44 & \cellcolor{gray3}70.67 & 52.45 & 57.94 & \cellcolor{gray2}87.32 & \cellcolor{gray2}87.00\\
    \midrule
    \multirow{6}{*}{\shortstack{C\# }}
    & Incoder & 48.75 & 48.97 & 53.59 & 56.97 & \cellcolor{gray2}82.05 & 33.17 & 64.66 & 65.12\\
    & Phi-1 & \cellcolor{gray1}\textbf{90.12} & 41.50 & \cellcolor{gray2}84.61 & \cellcolor{gray2}86.90 & 36.59 & 39.28 & \cellcolor{gray2}84.54 & \cellcolor{gray2}86.11\\
    & StarCoder & 62.09 & 55.50 & \cellcolor{gray3}72.55 & 67.69 & \cellcolor{gray3}71.59 & 27.34 & \cellcolor{gray3}73.39 & \cellcolor{gray3}73.62\\
    & CodeGen2 & \cellcolor{gray3}75.45 & 45.08 & \cellcolor{gray3}77.50 & \cellcolor{gray3}77.80 & 68.86 & 35.67 & \cellcolor{gray3}78.85 & \cellcolor{gray3}78.55\\
    & WizCoder & \cellcolor{gray3}76.73 & 36.49 & \cellcolor{gray3}75.41 & \cellcolor{gray3}76.48 & 54.82 & 33.22 & \cellcolor{gray3}76.45 & \cellcolor{gray3}79.04\\
    & CodeLlama & 62.96 & 59.45 & \cellcolor{gray3}72.53 & 67.66 & \cellcolor{gray3}71.32 & 35.06 & 66.94 & 68.34\\
    \cmidrule{2-10}
    & Average & 69.35 & 47.83 & \cellcolor{gray3}72.70 & \cellcolor{gray3}72.25 & 64.21 & 33.96 & \cellcolor{gray3}74.14 & \cellcolor{gray3}75.13\\
    \hline
    \multirow{6}{*}{\shortstack{C++}}
         & Incoder & 63.99 & 34.96 & 19.79 & 67.89 & 69.92 & 56.17 & \cellcolor{gray2}81.90 & \cellcolor{gray2}82.38\\
        & Phi-1 & \cellcolor{gray2}88.38 & 35.20 & 63.85 & \cellcolor{gray2}82.41 & 34.02 & 59.22 & \cellcolor{gray1}91.14 & \cellcolor{gray1}\textbf{92.10}\\
        & StarCoder & \cellcolor{gray3}75.17 & 44.86 & 56.75 & \cellcolor{gray3}74.79 & \cellcolor{gray3}70.28 & 54.83 & \cellcolor{gray2}89.56 & \cellcolor{gray2}89.35\\
        & CodeGen2 & \cellcolor{gray3}79.91 & 39.33 & 59.04 & \cellcolor{gray3}74.43 & 57.40 & 56.00 & \cellcolor{gray2}86.33 & \cellcolor{gray2}86.04\\
         & WizCoder & \cellcolor{gray2}83.06 & 31.19 & 61.39 & \cellcolor{gray2}83.11 & 56.86 & 57.46 & \cellcolor{gray1}90.52 & \cellcolor{gray1}91.37\\
        & CodeLlama & \cellcolor{gray3}76.00 & 48.37 & 59.64 & \cellcolor{gray3}77.39 & 67.68 & 56.51 & \cellcolor{gray2}86.08 & \cellcolor{gray2}86.65\\
        \cmidrule{2-10}
         & Average & \cellcolor{gray3}77.75 & 38.99 & 53.41 & \cellcolor{gray3}76.67 & 59.36 & 56.70 & \cellcolor{gray2}87.59 & \cellcolor{gray2}87.98\\
        \midrule
         \multirow{6}{*}{\shortstack{Go}}
        & Incoder & 41.91 & 53.58 & 22.65 & 43.97 & \cellcolor{gray3}71.24 & 29.42 & 54.83 & 54.47\\
        & Phi-1 & \cellcolor{gray3}\textbf{72.68} & 45.99 & 61.24 & 68.27 & 40.24 & 37.04 & 68.74 & \cellcolor{gray3}70.10\\
        & StarCoder & 50.95 & 60.61 & 52.13 & 49.62 & 47.88 & 32.19 & 53.33 & 53.59\\
        & CodeGen2 & 53.17 & 62.07 & 54.42 & 53.88 & 45.76 & 35.64 & 55.68 & 55.54\\
        & WizCoder & 64.71 & 60.73 & 48.72 & 57.91 & 45.15 & 32.09 & 62.42 & 64.21\\
        & CodeLlama & 47.84 & 65.36 & 53.63 & 49.98 & 49.25 & 32.54 & 53.00 & 54.40\\
        \cmidrule{2-10}
        & Average & 55.21 & 58.06 & 48.80 & 53.94 & 49.92 & 33.15 & 58.00 & 58.72\\
        \midrule
         \multirow{6}{*}{\shortstack{Java}}
         & Incoder & 54.61 & 44.73 & 21.18 & 61.72 & 67.74 & 42.03 & \cellcolor{gray3}75.11 & \cellcolor{gray3}74.65\\
        & Phi-1 & \cellcolor{gray2}85.44 & 40.40 & 62.55 & \cellcolor{gray3}79.07 & 28.80 & 49.80 & \cellcolor{gray2}84.69 & \cellcolor{gray2}\textbf{85.54}\\
        & StarCoder & 63.01 & 56.90 & 55.61 & 61.27 & 53.33 & 42.25 & \cellcolor{gray3}74.87 & \cellcolor{gray3}73.67 \\
        & CodeGen2 & \cellcolor{gray3}71.96 & 50.69 & 51.89 & 66.52 & 49.20 & 44.81 & \cellcolor{gray3}72.75 & \cellcolor{gray3}71.94 \\
        & WizCoder & 69.97 & 45.14 & 59.85 & 65.34 & 47.12 & 43.20 & 69.29 & \cellcolor{gray3}70.36\\
        & CodeLlama & 64.27 & 57.67 & 54.88 & 61.21 & 56.86 & 41.67 & 65.65 & 65.95 \\
        \cmidrule{2-10}
        & Average  & 68.21 & 49.26 & 50.99 & 65.85 & 50.51 & 43.96 & \cellcolor{gray3}73.73 & \cellcolor{gray3}75.23 \\
           \midrule
         \multirow{6}{*}{\shortstack{Python}}
         & Incoder & 47.33 & 51.23 & 44.73 & 50.34 & 57.62 & 41.02 & 60.91 & 58.52\\
        & Phi-1 & 60.03 & 53.48 & 48.46 & 56.93 & 34.82 & 45.08 & \textbf{63.89} & 62.64\\
        & StarCoder & 55.72 & 53.83 & 40.02 & 52.48 & 37.86 & 40.96 & 58.03 & 55.70 \\
        & CodeGen2 & 56.54 & 53.36 & 42.20 & 50.80 & 34.18 & 38.70 & 59.02 & 56.48 \\
        & WizCoder & 62.79 & 44.41 & 47.66 & 56.25 & 23.82 & 42.54 & 63.61 & 62.52\\
        & CodeLlama & 57.11 & 48.76 & 43.67 & 48.84 & 32.16 & 40.25 & 55.94 & 54.60 \\
        \cmidrule{2-10}
         & Average  & 56.59 & 50.84 & 44.46 & 52.61 & 36.74 & 41.43 & 60.23 & 58.41 \\
           \midrule
         \multirow{6}{*}{\shortstack{Ruby}}
         & Incoder & 52.41 & 44.31 & 46.83 & 51.24 & 61.11 & 39.17 & 55.49 & 54.16\\
        & Phi-1 & 67.47 & 39.48 & 63.86 & 65.58 & 54.34 & 44.45 & 66.32 & 65.64\\
        & StarCoder & 61.61 & 44.25 & 47.51 & 59.77 & 42.90 & 38.70 & 60.04 & 57.52 \\
        & CodeGen2 & 63.78 & 40.34 & 53.65 & 59.29 & 45.46 & 41.44 & 63.95 & 61.61 \\
        & WizCoder & \cellcolor{gray3}70.22 & 33.63 & 50.58 & 66.85 & 41.83 & 44.25 & \cellcolor{gray3}\textbf{73.44} & \cellcolor{gray3}71.65\\
        & CodeLlama & 61.63 & 42.48 & 57.81 & 65.62 & 58.55 & 39.15 & \cellcolor{gray3}73.17 & \cellcolor{gray3}72.05 \\
        \cmidrule{2-10}
         & Average  & 62.85 & 40.75 & 53.37 & 61.39 & 50.70 & 41.19 & 65.40 & 63.77 \\
           \midrule
         \multirow{6}{*}{\shortstack{Rust}}
        & Incoder & 68.41& 38.70& 69.87& \cellcolor{gray3}71.97& 23.03& \cellcolor{gray3}79.82& \cellcolor{gray3}71.55& \cellcolor{gray3}72.04\\
        & Phi-1 & \cellcolor{gray2}\textbf{89.72}& 20.69& \cellcolor{gray2}80.27& \cellcolor{gray2}88.28& 26.17& 61.87& \cellcolor{gray2}82.22& \cellcolor{gray2}83.63\\
        & StarCoder & 39.10& \cellcolor{gray3}73.59 & 49.26 & 42.53 & 23.63 & 55.23& 43.45 & 43.14 \\
        & CodeGen2 & 66.15 & 54.97 & 58.30 & 66.04 & 23.94 & 55.24 & 63.06 & 63.15 \\
        & WizCoder & 64.41 & 51.38 & 57.72 & 57.91 & 27.22& 36.61 & 55.29 & 55.96\\
        & CodeLlama & 57.79& 66.74 & 59.09 & 58.60 & 24.65 & 53.55 & 55.54 & 56.44 \\
        \cmidrule{2-10}
        & Average & 64.26 & 51.01 & 62.42 & 64.22 & 24.77 & 57.05 & 61.85 & 62.39 \\
        \midrule
\end{longtable}
}

\begin{table}[ht]
\footnotesize
    \centering
    \setlength{\tabcolsep}{0.7mm}
    \caption{The AUC of feature- and pre-training-based methods in terms of programming languages}
    \label{tab:performance-feature}
    \renewcommand\arraystretch{1.1}
    \setlength{\tabcolsep}{10pt}
    \begin{tabular}{lcc|c}
        \hline
         \multirow{2}{*}{\textbf{Language}} & \multicolumn{2}{c}{\textbf{Feature}} & \multicolumn{1}{c}{\textbf{Pre-training}}\\
        \cline{2-4}
        & Idialu (Python) & Bukhari (C) & GPTSniffer \\
        \hline
        
        C & / & \cellcolor{gray1}99.01 & \cellcolor{gray1}99.13  \\
        C\# & / & / & \cellcolor{gray1}98.15 \\
        C++ & / & / & \cellcolor{gray1}99.58 \\
        Go & / & / & \cellcolor{gray1}94.19 \\
        Java & / & / & \cellcolor{gray1}96.11 \\
        Python & \cellcolor{gray2}89.60 & / & \cellcolor{gray1}95.17 \\
        Ruby & / & / & \cellcolor{gray1}93.25 \\
        Rust & / & / & \cellcolor{gray1}98.64 \\
        \hline
    \end{tabular}
\end{table}

When comparing \per against the other two classes of methods, we select the best-performing method as representative\footnote{For succinctness, we adopt this measure throughout the paper}. 

\per does not perform as well as GPTSniffer, which is a pre-training-based method in all eight programming languages. 
The AUC of \per and GPTSniffer for each language are as follows: C (90.40\%--99.13\%), C\# (90.12\%--98.15\%), C++ (91.37\%--99.58\%), Go (72.68\%--94.19\%), Java (85.54\%--96.11\%), Python (63.89\%--95.17\%), Ruby (73.44\%--93.25\%), and Rust (89.72\%--98.64\%). 
The difference can be explained that GPTSniffer is a supervised method while \per is not. 
Moreover, \per cannot achieve the same performance as feature-based methods in Python and C: Python (63.89\%--89.60\%), C (90.40\%--99.01\%).

When comparing the AUC of \per between programming languages, we observe that \per performs best in C/C++ in which NPR and DetectCodeGPT achieve an average AUC of 85\%. 
However, NPR, DetectCodeGPT, and other methods cannot achieve an AUC of over 80\% in other languages. 
Taking C/C++ and Python as examples. 
C/C++ requires developers to implement many functions, which are more likely to be provided by third-party libraries in Python. 
In this case, C/C++ has more code stylometry than Python~\cite{caliskan2015anonymizing}. 
However, the design philosophy of Python encourages simplified code stylometry, making it easy for \llm to generate human-style code.

\begin{tcolorbox}[width=\linewidth, boxrule=1pt, sharp corners=all,
  left=2pt, right=2pt, top=2pt, bottom=2pt, colback=gray!8, colframe=black]
  $\textbf{RQ}_\textbf{1.1}$: 
  \per can not perform as well as the feature- and pre-training-based methods from the perspective of programming languages. Among eight languages, \per is only effective in C and C++. 
\end{tcolorbox}

\subsection{$\text{RQ}_\text{1.2}$: Degree of Difficulty}
\Cref{tab:diff-perplexity} and \Cref{tab:diff-feature} present the AUC of the three groups of methods from the perspective of the degree of difficulty.

\begin{table}[ht]
\footnotesize
    \centering
    \setlength{\tabcolsep}{0.9mm}
    \caption{The AUC of \per in terms of degrees of difficulty}
    \label{tab:diff-perplexity}
    \begin{tabular}{llccccc|ccc}
        \midrule
        \multirow{2}{*}{\textbf{Difficulty}} & \multirow{2}{*}{\textbf{Code LLM}} & \multicolumn{5}{c}{\textbf{without perturbation}} & \multicolumn{3}{c}{\textbf{with perturbation}} \\
        \cmidrule{3-10}
        & & log-p & Ent & Rank & LRank & LRR & DGPT & NPR & DCGPT \\
      \midrule
         \multirow{6}{*}{\shortstack{Easy}}
         & Incoder & 51.75 & 46.33 & 40.75 & 57.76 & 61.56 & 46.80 & 65.03 & 64.46 \\
        & Phi-1 & 69.63 & 44.47 & 57.08 & 69.34 & 38.90 & 57.84 & \cellcolor{gray3}\textbf{75.41} & \cellcolor{gray3}75.09 \\
        & StarCoder & 56.23 & 52.40 & 49.72 & 56.86 & 50.90 & 46.27 & 64.13 & 62.79 \\
        & CodeGen2 & 62.19 & 46.78 & 53.82 & 59.32 & 49.77 & 45.71 & 65.50 & 64.02 \\
        & WizCoder & 67.84 & 41.60 & 55.04 & 59.52 & 39.82 & 47.72 & 64.72 & 64.31 \\
        & CodeLlama & 59.84 & 49.69 & 53.81 & 58.15 & 52.16 & 45.62 & 63.60 & 62.91 \\
        \cmidrule{2-10}
        & Average & 61.25 & 46.88 & 51.70 & 60.16 & 48.85 & 48.33 & 66.40 & 65.60 \\
        \midrule
        \multirow{6}{*}{\shortstack{Middle}}
        & Incoder & 53.30 & 45.55 & 41.57 & 57.04 & 66.77 & 42.51 & 66.02 & 65.56 \\
        & Phi-1 & \cellcolor{gray3}76.54 & 39.33 & 65.22 & \cellcolor{gray3}71.44 & 42.02 & 46.18 & \cellcolor{gray3}76.29 & \cellcolor{gray3}\textbf{77.00} \\
        & StarCoder & 55.35 & 56.54 & 52.41 & 54.32 & 49.97 & 43.12 & 61.36 & 60.56 \\
        & CodeGen2 & 62.77 & 49.94 & 53.28 & 63.96 & 53.03 & 46.77 & 69.02 & 68.22 \\
        & WizCoder & 66.38 & 43.22 & 56.73 & 62.90 & 43.73 & 43.51 & 68.18 & 68.69 \\
        & CodeLlama & 57.67 & 54.05 & 52.91 & 60.93 & 55.40 & 41.20 & 66.08 & 65.97 \\
        \cmidrule{2-10}
        & Average & 62.00 & 48.11 & 53.69 & 61.76 & 51.82 & 43.88 & 67.82 & 67.67 \\
        \midrule
        \multirow{6}{*}{\shortstack{Hard}}
        & Incoder & 54.53 & 44.44 & 37.73 & 64.05 & \cellcolor{gray3}\textbf{76.48} & 41.64 & \cellcolor{gray3}70.27 & \cellcolor{gray3}70.92 \\
        & Phi-1 & \cellcolor{gray3}75.92 & 40.94 & \cellcolor{gray3}70.32 & \cellcolor{gray3}72.25 & 46.92 & 41.63 & \cellcolor{gray3}74.69 & \cellcolor{gray3}76.07 \\
        & StarCoder & 56.30 & 57.90 & 58.33 & 54.99 & 53.88 & 34.26 & 62.48 & 62.47 \\
        & CodeGen2 & 65.25 & 50.16 & 61.54 & 67.26 & 57.51 & 39.70 & \cellcolor{gray3}71.23 & \cellcolor{gray3}71.21 \\
        & WizCoder & 66.28 & 41.05 & 63.58 & 66.32 & 49.05 & 36.55 & 69.21 & \cellcolor{gray3}70.53 \\
        & CodeLlama & 62.97 & 54.87 & 65.99 & 62.32 & 60.48 & 35.16 & 65.82 & 66.67 \\
        \cmidrule{2-10}
        & Average & 63.54 & 48.23 & 59.58 & 64.53 & 57.39 & 38.16 & 68.95 & 69.64 \\

        \midrule
    \end{tabular}
\end{table}

\begin{table}[ht]
\footnotesize
    \centering
    \setlength{\tabcolsep}{0.7mm}
    \caption{The AUC of feature- and pre-training-based methods in terms of degrees of difficulty}
    \label{tab:diff-feature}
    \renewcommand\arraystretch{1.1}
    \setlength{\tabcolsep}{10pt}
    \begin{tabular}{lcc|c}
        \midrule
         \multirow{2}{*}{\textbf{Difficulty}} & \multicolumn{2}{c}{\textbf{Feature}} & \multicolumn{1}{c}{\textbf{Pre-training}}\\
        \cline{2-4}
        & Idialu (Python) & Bukhari (C) & GPTSniffer \\
        \hline
        
        Easy & \cellcolor{gray2}85.37 & \cellcolor{gray1}96.40 & \cellcolor{gray1}90.57  \\
        Middle & \cellcolor{gray2}89.60 & \cellcolor{gray1}98.98 & \cellcolor{gray1}94.32 \\
        Hard & \cellcolor{gray1}93.18 & \cellcolor{gray1}98.25 & \cellcolor{gray1}94.33 \\
        \hline
    \end{tabular}
\end{table}

\per exhibits lower performance compared with the feature- and pre-training-based methods in terms of all three degrees of difficulty. 
The differences between \per and pre-training-based methods for each level of difficulty are as follows: \emph{easy} (75.41\%--90.57\%), \emph{middle} (77.00\%--94.32\%), and \emph{hard} (76.48\%--94.33\%).
When compared with feature-based methods, \per is less effective than Idialu: \emph{easy} (75.41\%--85.37\%), \emph{middle} (77.00\%--89.60\%), and \emph{hard} (76.48\%--93.18\%), and Bukhari: \emph{easy} (75.41\%--96.40\%), \emph{middle} (77.00\%--98.98\%), and \emph{hard} (76.48\%--98.25\%). 
The discrepancy can be explained that feature- and pre-training-based methods are based on supervised training, while \per not. 

When it comes to the AUC of \per from degrees of difficulty, we find that \per does not perform well in all three degrees of difficulty, but slightly improves from the easy to the hard level. 
To be specific, log p(x) and Log Rank are the best-performing methods in \per without perturbation, achieving the AUC of 61.25\% and 60.10\% at the easy level, 62.00\%, and 61.76\% at the middle level, and 63.54\% and 64.53\% at the hard level, respectively. 
Similarly, the best-performing methods in \per with perturbation, \ie NPR and DetectCodeGPT, achieve the AUC of 66.40\% and 65.60\% at the easy level, 67.82\% and 67.67\% at the middle level, and 68.95\% and 69.64\% at the hard level, respectively. 
Since the solutions to difficult puzzles have complex logic and customized implementations, they have more code stylometry than solutions to easy puzzles.

\begin{tcolorbox}[width=\linewidth, boxrule=1pt, sharp corners=all,
  left=2pt, right=2pt, top=2pt, bottom=2pt, colback=gray!8, colframe=black]
  $\textbf{RQ}_\textbf{1.2}$: \per perform not well in all three degrees of difficulty, even though the performance slightly improves from the easy to the hard level. The feature- and pre-training-based methods perform well at all three degrees of difficulty.
\end{tcolorbox}

\subsection{$\text{RQ}_\text{1.3}$: Scale of Solution}
\Cref{tab:scale-perplexity} and \Cref{tab:scale-feature} show the AUC of three classes of methods in terms of the scale of solutions. Note that in \Cref{tab:scale-feature}, we provide results only in the small and middle scales for Idialu due to no large-scale code solution in Python.

\begin{table}[ht]
\footnotesize
    \centering
    \setlength{\tabcolsep}{0.9mm}
    \caption{The AUC of \per in terms of scales of solution}
    \label{tab:scale-perplexity}
    \begin{tabular}{llccccc|ccc}
        \midrule
        \multirow{2}{*}{\textbf{Scale}} & \multirow{2}{*}{\textbf{Code LLM}} & \multicolumn{5}{c}{\textbf{without perturbation}} & \multicolumn{3}{c}{\textbf{with perturbation}} \\
        \cmidrule{3-10}
        & & log-p & Ent & Rank & L-Rank & LRR & DGPT & NPR & DCGPT \\
        \midrule
         \multirow{6}{*}{\shortstack{Small}}
        & Incoder & 52.94 & 45.41 & 47.20 & 52.79 & 53.91 & 47.55 & 58.46 & 57.24 \\
        & Phi-1 & 65.07 & 42.43 & 53.11 & 62.33 & 42.86 & 60.14 & \textbf{69.75} & 68.63 \\
        & StarCoder & 61.72 & 43.41 & 49.76 & 53.42 & 41.05 & 54.93 & 59.29 & 56.89 \\
        & CodeGen2 & 60.65 & 43.23 & 50.69 & 55.45 & 44.03 & 47.30 & 63.75 & 61.25 \\
        & WizCoder & 66.42 & 38.99 & 53.71 & 63.48 & 40.26 & 52.51 & 69.47 & 68.01 \\
        & CodeLlama & 61.68 & 40.38 & 54.39 & 60.45 & 50.67 & 43.50 & 67.95 & 66.35 \\
        \cmidrule{2-10}
        & Average & 61.41 & 42.31 & 51.48 & 57.99 & 45.46 & 50.99 & 64.78 & 63.06 \\
        \midrule
        \multirow{6}{*}{\shortstack{Middle}}
        & Incoder & 57.47 & 43.11 & 39.15 & 61.29 & 63.56 & 57.28 & \cellcolor{gray3}72.93 & \cellcolor{gray3}72.04 \\
        & Phi-1 & \cellcolor{gray3}75.35 & 38.13 & 55.44 & 67.33 & 35.52 & 59.11 & \cellcolor{gray3}\textbf{77.42} & \cellcolor{gray3}77.26 \\
        & StarCoder & 63.84 & 47.49 & 54.01 & 59.42 & 51.61 & 56.54 & \cellcolor{gray3}70.22 & 69.36 \\
        & CodeGen2 & 68.56 & 41.39 & 51.73 & 63.92 & 44.98 & 56.55 & \cellcolor{gray3}72.55 & \cellcolor{gray3}71.29 \\
        & WizCoder & \cellcolor{gray3}72.77 & 34.01 & 52.63 & 63.36 & 38.64 & 57.51 & \cellcolor{gray3}73.97 & \cellcolor{gray3}73.64 \\
        & CodeLlama & 63.85 & 46.76 & 51.77 & 62.86 & 50.66 & 54.37 & \cellcolor{gray3}70.38 & 69.56 \\
        \cmidrule{2-10}
        & Average & 66.97 & 41.82 & 50.79 & 63.03 & 47.50 & 56.89 & \cellcolor{gray3}72.91 & \cellcolor{gray3}72.19 \\
        \midrule
        \multirow{6}{*}{\shortstack{Large}}
        & Incoder & \cellcolor{gray3}72.17 & 29.05 & 47.88 & \cellcolor{gray3}75.28 & \cellcolor{gray3}77.86 & 48.97 & \cellcolor{gray3}77.12 & \cellcolor{gray3}78.39 \\
        & Phi-1 & \cellcolor{gray2}\textbf{87.81} & 21.03 & \cellcolor{gray3}73.76 & \cellcolor{gray2}86.00 & 58.52 & 50.81 & \cellcolor{gray2}82.22 & \cellcolor{gray2}83.94 \\
        & StarCoder & 67.85 & 46.41 & 61.18 & 67.59 & 58.67 & 45.13 & 68.30 & 69.46 \\
        & CodeGen2 & \cellcolor{gray3}78.21 & 35.07 & 65.32 & \cellcolor{gray3}77.19 & 61.70 & 43.76 & \cellcolor{gray3}75.46 & \cellcolor{gray3}76.62 \\
        & WizCoder & \cellcolor{gray2}82.75 & 21.61 & \cellcolor{gray3}71.10 & \cellcolor{gray2}81.27 & 57.75 & 47.70 & \cellcolor{gray2}81.04 & \cellcolor{gray2}82.24 \\
        & CodeLlama & \cellcolor{gray3}76.40 & 42.48 & 69.43 & \cellcolor{gray3}76.34 & 64.85 & 45.54 & \cellcolor{gray3}76.70 & \cellcolor{gray3}77.94 \\
        \cmidrule{2-10}
        & Average & \cellcolor{gray3}77.53 & 32.61 & 64.78 & \cellcolor{gray3}77.28 & 63.22 & 46.99 & \cellcolor{gray3}76.81 & \cellcolor{gray3}78.10 \\
        \midrule
    \end{tabular}
\end{table}

\begin{table}[ht]
\footnotesize
    \centering
    \setlength{\tabcolsep}{0.7mm}
    \caption{The AUC of feature- and pre-training-based methods in terms of scales of solution}
    \label{tab:scale-feature}
    \renewcommand\arraystretch{1.1}
    \setlength{\tabcolsep}{10pt}
    \begin{tabular}{lcc|c}
        \hline
         \multirow{2}{*}{\textbf{Scale}} & \multicolumn{2}{c}{\textbf{Feature}} & \multicolumn{1}{c}{\textbf{Pre-training}}\\
        \cline{2-4}
        & Idialu (Python) & Bukhari (C) & GPTSniffer \\
        \hline
        
        Small & \cellcolor{gray2}85.87 & \cellcolor{gray1}91.20 & \cellcolor{gray2}82.74  \\
        Middle & \cellcolor{gray1}92.58 & \cellcolor{gray1}98.26 & \cellcolor{gray1}96.10 \\
        Large & / & \cellcolor{gray1}96.67 & \cellcolor{gray1}94.19 \\
        \hline
    \end{tabular}
\end{table}

\per does not perform as well as the pre-training-based method, \ie GPTSniffer in all three scales. 
The AUC of \per and GPTSniffer for each scale is: small (69.75\%--82.74\%), middle (77.42\%--96.10\%), and large (87.81\%--94.19\%). 
For the featured-based method, \per is less effective than Idialu: small (69.75\%--85.87\%), middle (77.42\%--92.58\%), and Bukhari: small (69.75\%--91.20\%), middle (77.42\%--98.26\%), and large (87.81\%--96.67\%). 
These differences are mainly because feature- and pre-training-based methods are supervised, while \per is a collection of heuristic methods.

When comparing the methods in \per regarding different scales, we observe that those well-performed methods improve from the small to the large scale. For example, log p(x) is the best-performing method in \per without perturbation, with an AUC of 61.41\% for small, 66.97\% for middle, and 77.53\% for large. Similarly, the best-performing methods in \per with perturbation, \ie NPR and DetectCodeGPT, achieve an AUC of 64.87\% and 63.06\% for small, 72.91\% and 72.19\% for middle, and 76.81\% and 78.10\% for large, respectively. These results can be explained that large-scale code snippets have more diverse and detailed implementations than others, which makes \hac more distinct from \lgc.

\begin{tcolorbox}[width=\linewidth, boxrule=1pt, sharp corners=all,
  left=2pt, right=2pt, top=2pt, bottom=2pt, colback=gray!8, colframe=black]
  $\textbf{RQ}_\textbf{1.3}$: \per performs less effectively compared with feature- and pre-training-based methods in all the three scales of code solutions. \per only exhibits acceptable performance in the large-scale code solution.
\end{tcolorbox}

\subsection{$\text{RQ}_\text{2}$: Detection Speed}

\begin{table}[ht]
\footnotesize
    \centering
    \setlength{\tabcolsep}{0.9mm}
    \caption{The detection speed of \per in terms of \llm (seconds per sample)}
    \label{tab:effciency-perplexity}
    \begin{tabular}{llccccc|ccc}
        \midrule
        \multirow{2}{*}{\textbf{Model}} & \multirow{2}{*}{\textbf{Code LLM}} & \multicolumn{5}{c}{\textbf{without perturbation}} & \multicolumn{3}{c}{\textbf{with perturbation}} \\
        \cmidrule{3-10}
        & & log-p & Ent & Rank & L-Rank & LRR & DGPT & NPR & DCGPT \\
      \midrule
      \multirow{6}{*}{\shortstack{GPT-3.5}}
      & Incoder & \cellcolor{gray3}0.122 & \cellcolor{gray2}0.092 & \cellcolor{gray3}0.111 & \cellcolor{gray3}0.128 & \cellcolor{gray3}0.235 & 143.5 & 5.306 & 5.306 \\
      & Phi-1 & \cellcolor{gray3}0.148 & \cellcolor{gray3}0.118 & \cellcolor{gray3}0.137 & \cellcolor{gray3}0.151 & \cellcolor{gray3}0.282 & 170.7 & 6.307 & 6.308 \\
      & StarCoder & \cellcolor{gray3}0.273 & \cellcolor{gray3}0.231 & \cellcolor{gray3}0.248 & \cellcolor{gray3}0.267 & \cellcolor{gray3}0.508 & 152.0 & 11.12 & 11.12 \\
      & CodeGen2 & \cellcolor{gray3}0.328 & \cellcolor{gray3}0.281 & \cellcolor{gray3}0.306 & \cellcolor{gray3}0.319 & \cellcolor{gray3}0.606 & 148.3 & 13.27 & 13.28 \\
      & WizCoder & \cellcolor{gray3}0.640 & \cellcolor{gray3}0.561 & \cellcolor{gray3}0.577 & \cellcolor{gray3}0.656 & 1.265 & 156.1 & 30.27 & 30.27 \\
      & CodeLlama & \cellcolor{gray3}0.106 & \cellcolor{gray2}\textbf{0.076} & \cellcolor{gray2}0.096 & \cellcolor{gray3}0.135 & \cellcolor{gray3}0.240 & 146.7 & 5.545 & 5.545 \\
      \cmidrule{2-10}
      & Average & \cellcolor{gray3}0.270 & \cellcolor{gray3}0.227 & \cellcolor{gray3}0.246 & \cellcolor{gray3}0.276 & \cellcolor{gray3}0.523 & 152.9 & 11.97 & 11.97\\
      \midrule
      \multirow{6}{*}{\shortstack{GPT-4o}}
      & Incoder & \cellcolor{gray3}0.127 & \cellcolor{gray2}0.099 & \cellcolor{gray3}0.118 & \cellcolor{gray3}0.134 & \cellcolor{gray3}0.248 & 143.1 & 5.619 & 5.620 \\
      & Phi-1 & \cellcolor{gray3}0.157 & \cellcolor{gray3}0.126 & \cellcolor{gray3}0.146 & \cellcolor{gray3}0.163 & \cellcolor{gray3}0.305 & 143.5 & 6.738 & 6.740 \\
      & StarCoder & \cellcolor{gray3}0.292 & \cellcolor{gray3}0.249 & \cellcolor{gray3}0.265 & \cellcolor{gray3}0.289 & \cellcolor{gray3}0.553 & 148.9 & 11.97 & 11.97 \\
      & CodeGen2 & \cellcolor{gray3}0.360 & \cellcolor{gray3}0.303 & \cellcolor{gray3}0.321 & \cellcolor{gray3}0.350 & \cellcolor{gray3}0.662 & 151.0 & 14.29 & 14.29 \\
      & WizCoder & \cellcolor{gray3}0.691 & \cellcolor{gray3}0.604 & \cellcolor{gray3}0.623 & \cellcolor{gray3}0.708 & 1.379 & 155.2 & 33.08 & 33.08 \\
      & CodeLlama & \cellcolor{gray3}0.108 & \cellcolor{gray2}\textbf{0.080} & \cellcolor{gray2}0.100 & \cellcolor{gray3}0.139 & \cellcolor{gray3}0.248 & 143.9 & 5.800 & 5.800 \\
      \cmidrule{2-10}
      & Average & \cellcolor{gray3}0.289 & \cellcolor{gray3}0.244 & \cellcolor{gray3}0.262 & \cellcolor{gray3}0.297 & \cellcolor{gray3}0.566 & 147.6 & 12.92 & 12.92 \\
      \midrule
      \multirow{6}{*}{\shortstack{Gemini-1.0}}
      & Incoder & \cellcolor{gray3}0.122 & \cellcolor{gray2}0.094 & \cellcolor{gray3}0.113 & \cellcolor{gray3}0.130 & \cellcolor{gray3}0.238 & 173.0 & 5.370 & 5.370 \\
      & Phi-1 & \cellcolor{gray3}0.150 & \cellcolor{gray3}0.121 & \cellcolor{gray3}0.140 & \cellcolor{gray3}0.151 & \cellcolor{gray3}0.285 & 146.1 & 6.400 & 6.400 \\
      & StarCoder & \cellcolor{gray3}0.278 & \cellcolor{gray3}0.237 & \cellcolor{gray3}0.255 & \cellcolor{gray3}0.267 & \cellcolor{gray3}0.509 & 169.9 & 11.14 & 11.14 \\
      & CodeGen2 & \cellcolor{gray3}0.336 & \cellcolor{gray3}0.287 & \cellcolor{gray3}0.306 & \cellcolor{gray3}0.328 & \cellcolor{gray3}0.618 & 146.8 & 13.37 & 13.37 \\
      & WizCoder & \cellcolor{gray3}0.654 & \cellcolor{gray3}0.575 & \cellcolor{gray3}0.591 & \cellcolor{gray3}0.672 & 1.288 & 161.0 & 30.31 & 30.31 \\
      & CodeLlama & \cellcolor{gray3}0.105 & \cellcolor{gray2}\textbf{0.078} & \cellcolor{gray2}0.098 & \cellcolor{gray3}0.132 & \cellcolor{gray3}0.237 & 143.8 & 5.569 & 5.570 \\
      \cmidrule{2-10}
      & Average & \cellcolor{gray3}0.274 & \cellcolor{gray3}0.232 & \cellcolor{gray3}0.251 & \cellcolor{gray3}0.280 & \cellcolor{gray3}0.529 & 156.8 & 12.03 & 12.03 \\
      \midrule
      \multirow{6}{*}{\shortstack{Llama-3.1}}
      & Incoder & \cellcolor{gray3}0.124 & \cellcolor{gray2}0.096 & \cellcolor{gray3}0.114 & \cellcolor{gray3}0.139 & \cellcolor{gray3}0.248 & 172.0 & 5.406 & 5.407 \\
      & Phi-1 & \cellcolor{gray3}0.153 & \cellcolor{gray3}0.123 & \cellcolor{gray3}0.141 & \cellcolor{gray3}0.155 & \cellcolor{gray3}0.289 & 165.7 & 6.479 & 6.481 \\
      & StarCoder & \cellcolor{gray3}0.283 & \cellcolor{gray3}0.241 & \cellcolor{gray3}0.258 & \cellcolor{gray3}0.273 & \cellcolor{gray3}0.522 & 162.7 & 11.38 & 11.38 \\
      & CodeGen2 & \cellcolor{gray3}0.342 & \cellcolor{gray3}0.292 & \cellcolor{gray3}0.312 & \cellcolor{gray3}0.328 & \cellcolor{gray3}0.622 & 150.8 & 13.58 & 13.58 \\
      & WizCoder & \cellcolor{gray3}0.670 & \cellcolor{gray3}0.586 & \cellcolor{gray3}0.603 & \cellcolor{gray3}0.425 & \cellcolor{gray3}0.801 & 223.7 & 16.39 & 16.39 \\
      & CodeLlama & \cellcolor{gray3}0.106 & \cellcolor{gray2}\textbf{0.079} & \cellcolor{gray2}0.099 & \cellcolor{gray3}0.124 & \cellcolor{gray3}0.217 & 142.6 & 4.879 & 4.879 \\
      \cmidrule{2-10}
      & Average & \cellcolor{gray3}0.280 & \cellcolor{gray3}0.236 &\cellcolor{gray3}0.255 & \cellcolor{gray3}0.241 & \cellcolor{gray3}0.450 & 169.6 & 9.685 & 9.685 \\
    
        \midrule
    \end{tabular}
\end{table}

\begin{table}[ht]
\footnotesize
    \tabcolsep=0.1cm
    \centering
    \caption{The detection speed of feature- and pre-training-based methods in terms of LLMs (seconds per sample)}
    \label{tab:efficiency-feature}
    \renewcommand\arraystretch{1.1}
    \setlength{\tabcolsep}{10pt}
    \begin{tabular}{lcc|c}
        \hline
         \multirow{2}{*}{\textbf{LLM}} & \multicolumn{2}{c}{\textbf{Feature}} & \multicolumn{1}{c}{\textbf{Pre-training}}\\
        \cline{2-4}
        & Idialu (Python) & Bukhari (C) & GPTSniffer \\
        \hline
        GPT-3.5 & \cellcolor{gray2}0.038 & 4.352 &\cellcolor{gray2}0.014  \\
        GPT-4o & \cellcolor{gray2}0.042 & 4.312 & \cellcolor{gray1}0.010 \\
        Gemini-1.0 & \cellcolor{gray2}0.040 & 4.504 & \cellcolor{gray1}0.010 \\
        Llama-3.1 & \cellcolor{gray2}0.040 & 4.794 & \cellcolor{gray1}0.010 \\
        \hline
    \end{tabular}
\end{table}

\Cref{tab:effciency-perplexity} and \Cref{tab:efficiency-feature} present the speed of the three classes of methods in detecting code generated by four \llm. For presentation clarity, we highlight notable results in blue (sec/sample: \colorbox{gray1}{$0-0.01$};\colorbox{gray2}{$0.01-0.1$};\colorbox{gray3}{$0.1-1$})

\per is not as fast as feature- and pre-training-based methods on all four LLMs. 
The AUC of \per and the pre-training-based method for each LLM are as follows: GPT-3.5 (0.076--0.014), GPT-4o (0.080--0.010), Gemini-1.0 (0.078--0.010), and Llama-3.1 (0.079--0.010). 
\per calculates the perplexity of each token (a code snippet consists of a set of tokens) using another LLM; therefore, it is very time-consuming.
When compared with the feature-based method, \per is slower than Idialu: GPT-3.5 (0.076--0.038), GPT-4o (0.080--0.042), Gemini-1.0 (0.078--0.040), and Llama-3.1 (0.079--0.040).
However, \per is faster than Bukhari: GPT-3.5 (0.076--4.352), GPT-4o (0.080--4.312), Gemini-1.0 (0.078--4.504), and Llama-3.1 (0.079--4.794). 
Since Bukhari extracts n-gram features from abstract syntactic trees, it has a slow detection speed.

When comparing the detection speed of methods in \per, we observe that methods without perturbation are much faster than methods with perturbation. 
The speed of the worst-performing method without perturbation is less than 2 seconds per sample: GPT-3.5 (1.265), GPT-4o (1.379), Gemini-1.0 (1.288), and Llama-3.1 (0.801). 
On the contrary, the speed of the best-performing method with perturbation is more than 4 seconds per sample: GPT-3.5 (5.306), GPT-4o (5.619), Gemini-1.0 (5.370), and Llama-3.1 (4.879). 
\per with perturbations introduces variations into the original sample to generate a list of perturbed samples and calculate attributes such as log-rank of two types of samples, 
whereas \per without perturbations only calculates attributes of the original sample.
Therefore, \per without perturbations performs much faster than \per with perturbations.
When comparing \per methods with perturbation, NPR and DetectCodeGPT have a faster detection speed than DetectGPT. 
The average speed of NPR and DetectCodeGPT is around 10 seconds per sample: GPT-3.5 (11.97--11.97), GPT-4o (12.92--12.92), Gemini-1.0 (12.03--12.03), and Llama-3.1 (9.685--9.685). 
NPR and DetectCodeGPT achieve very close performance because DetectCodeGPT is based on NPR. 
However, the average speed of DetectGPT is more than 100 seconds per sample: GPT-3.5 (152.9), GPT-4o (147.6), Gemini-1.0 (156.8), and Llama-3.1 (169.6). 
The perturbation method of DetectGPT is complex, \ie masking and completing tokens by another LLM. The perturbation method of DetectCodeGPT is simple, \ie adding spaces and new lines to the original tokens.

\begin{tcolorbox}[width=\linewidth, boxrule=1pt, sharp corners=all,
  left=2pt, right=2pt, top=2pt, bottom=2pt, colback=gray!8, colframe=black]
  $\textbf{RQ}_\textbf{2}$: \per exhibits a lower detection speed than the feature- and pre-training-based methods. The detection speed of \per without perturbation is faster than \per with perturbation. 
\end{tcolorbox}

\subsection{$\text{RQ}_\text{3}$: Generalization Capability}
\Cref{tab:gen-perplexity} and \Cref{tab:gen-feature} present the AUC of \per and the other two classes of methods in detecting \lgc generated by four \llm.
The LLMs listed in the two tables are used to generate \lgc.
The code LLMs listed in \Cref{tab:gen-perplexity} are used to calculate perplexity values.
For presentation clarity, we highlight notable results in blue (AUC:\colorbox{gray1}{$90\%-100\%$};\colorbox{gray2}{$80\%-90\%$};\colorbox{gray3}{$70\%-80\%$}).

\begin{table}[ht]
\footnotesize
    \centering
    \setlength{\tabcolsep}{0.9mm}
    \caption{The AUC of \per in terms of \llm}
    \label{tab:gen-perplexity}
    \begin{tabular}{llccccc|ccc}
        \midrule
        \multirow{2}{*}{\textbf{LLM}} & \multirow{2}{*}{\textbf{Code LLM}} & \multicolumn{5}{c}{\textbf{without perturbation}} & \multicolumn{3}{c}{\textbf{with perturbation}} \\
        \cmidrule{3-10}
        & & log-p & Ent & Rank & L-Rank & LRR & DGPT & NPR & DCGPT \\
      \midrule
       \multirow{6}{*}{\shortstack{GPT-3.5}}
        & Incoder & 51.89 & 46.86 & 39.93 & 63.01 & \cellcolor{gray3}72.28 & 49.68 & \cellcolor{gray3}73.10 & \cellcolor{gray3}72.78 \\
        & Phi-1 & \cellcolor{gray3}75.44 & 41.93 & 66.19 & \cellcolor{gray3}75.87 & 45.27 & 52.79 & \cellcolor{gray2}81.10 & \cellcolor{gray2}\textbf{81.64} \\
        & StarCoder & 56.14 & 54.82 & 55.86 & 61.09 & 57.55 & 47.53 & \cellcolor{gray3}70.86 & \cellcolor{gray3}70.03 \\
        & CodeGen2 & 63.27 & 49.26 & 58.56 & 66.84 & 56.52 & 48.27 & \cellcolor{gray3}73.31 & \cellcolor{gray3}72.40 \\
        & WizCoder & 66.34 & 42.77 & 60.98 & 68.54 & 50.25 & \cellcolor{gray3}72.24 & \cellcolor{gray3}74.26 & \cellcolor{gray3}74.37 \\
        & CodeLlama & 59.41 & 53.78 & 60.53 & 65.79 & 60.39 & 41.48 & \cellcolor{gray3}71.52 & \cellcolor{gray3}71.28 \\
        \cmidrule{2-10}
        & Average & 62.08 & 48.24 & 57.01 & 66.86 & 57.04 & 52.00 & \cellcolor{gray3}74.03 & \cellcolor{gray3}73.75 \\
        \midrule
        \multirow{6}{*}{\shortstack{GPT-4o}}
        & Incoder & 55.29 & 44.88 & 44.15 & 63.16 & 63.12 & 54.10 & \cellcolor{gray3}74.89 & \cellcolor{gray3}74.71 \\
        & Phi-1 & \cellcolor{gray3}72.89 & 38.02 & 63.79 & \cellcolor{gray3}72.56 & 44.10 & 58.14 & \cellcolor{gray2}80.52 & \cellcolor{gray2}\textbf{80.88} \\
        & StarCoder & 56.04 & 52.69 & 53.07 & 58.90 & 53.02 & 51.63 & \cellcolor{gray3}71.25 & \cellcolor{gray3}70.44 \\
        & CodeGen2 & 62.52 & 48.15 & 54.12 & 63.91 & 51.03 & 53.03 & \cellcolor{gray3}72.10 & \cellcolor{gray3}71.55 \\
        & WizCoder & 65.54 & 40.62 & 57.90 & 65.91 & 45.73 & 66.75 & \cellcolor{gray3}72.95 & \cellcolor{gray3}73.00 \\
        & CodeLlama & 60.70 & 50.93 & 56.88 & 64.68 & 54.86 & 48.01 & \cellcolor{gray3}71.13 & \cellcolor{gray3}70.93 \\
        \cmidrule{2-10}
        & Average & 62.16 & 45.88 & 54.98 & 64.85 & 51.98 & 55.28 & \cellcolor{gray3}73.81 & \cellcolor{gray3}73.59 \\
        \midrule
        \multirow{6}{*}{\shortstack{Gemini-1.0}}
        & Incoder & 54.53 & 45.43 & 42.55 & 63.70 & \cellcolor{gray3}76.00 & 51.37 & \cellcolor{gray3}73.74 & \cellcolor{gray3}73.17 \\
        & Phi-1 & \cellcolor{gray3}74.55 & 42.25 & \cellcolor{gray3}71.10 & \cellcolor{gray3}74.30 & 51.08 & 54.88 & \cellcolor{gray2}\textbf{81.78} & \cellcolor{gray2}81.66 \\
        & StarCoder & 60.58 & 53.99 & 64.18 & 64.75 & 66.84 & 49.24 & \cellcolor{gray3}73.36 & \cellcolor{gray3}72.26 \\
        & CodeGen2 & 67.39 & 48.34 & 65.84 & 69.68 & 65.25 & 50.68 & \cellcolor{gray3}76.23 & \cellcolor{gray3}74.98 \\
        & WizCoder & 69.21 & 44.70 & 64.43 & 67.99 & 50.52 & 58.07 & \cellcolor{gray3}74.36 & \cellcolor{gray3}74.15 \\
        & CodeLlama & 63.76 & 53.00 & 68.01 & 66.36 & 68.45 & 43.83 & \cellcolor{gray3}73.49 & \cellcolor{gray3}73.21 \\
        \cmidrule{2-10}
        & Average & 65.00 & 47.95 & 62.68 & 67.80 & 63.02 & 51.35 & \cellcolor{gray3}75.49 & \cellcolor{gray3}74.90 \\
        \midrule
        \multirow{6}{*}{\shortstack{Llama-3.1}}
        & Incoder & 57.00 & 42.88 & 43.15 & 65.10 & \cellcolor{gray3}74.82 & 50.84 & \cellcolor{gray3}73.20 & \cellcolor{gray3}73.14 \\
        & Phi-1 & \cellcolor{gray3}74.45 & 41.33 & 68.64 & \cellcolor{gray3}73.47 & 46.39 & 54.96 & \cellcolor{gray3}79.42 & \cellcolor{gray3}\textbf{79.55} \\
        & StarCoder & 66.20 & 47.63 & 67.67 & \cellcolor{gray3}70.29 & \cellcolor{gray3}72.05 & 51.47 & \cellcolor{gray3}75.56 & \cellcolor{gray3}75.23 \\
        & CodeGen2 & \cellcolor{gray3}71.42 & 42.79 & 66.51 & \cellcolor{gray3}73.05 & 66.92 & 49.79 & \cellcolor{gray3}77.42 & \cellcolor{gray3}76.95 \\
        & WizCoder & \cellcolor{gray3}74.69 & 38.59 & 68.99 & \cellcolor{gray3}76.39 & 56.50 & 67.27 & \cellcolor{gray3}79.08 & \cellcolor{gray3}78.74 \\
        & CodeLlama & 69.59 & 47.23 & \cellcolor{gray3}71.41 & \cellcolor{gray3}73.78 & \cellcolor{gray3}72.37 & 52.15 & \cellcolor{gray3}77.52 & \cellcolor{gray3}77.69 \\
        \cmidrule{2-10}
        & Average & 68.89 & 43.41 & 64.39 & \cellcolor{gray3}72.01 & 64.84 & 54.41 & \cellcolor{gray3}77.03 & \cellcolor{gray3}76.88 \\
        \midrule
    \end{tabular}
\end{table}

\begin{table}[ht]
\footnotesize
    \centering
    \setlength{\tabcolsep}{0.7mm}
    \caption{The AUC of feature- and pre-training-based methods in terms of \llm}
    \label{tab:gen-feature}
    \renewcommand\arraystretch{1.1}
    \setlength{\tabcolsep}{10pt}
    \begin{tabular}{lcc|c}
        \hline
         \multirow{2}{*}{\textbf{LLM}} & \multicolumn{2}{c}{\textbf{Feature}} & \multicolumn{1}{c}{\textbf{Pre-training}}\\
        \cline{2-4}
        & Idialu (Python) & Bukhari (C) & GPTSniffer \\
        \hline
        
        GPT-3.5 & \cellcolor{gray1}91.43 & \cellcolor{gray1}98.63 & \cellcolor{gray1}92.21  \\
        GPT-4o & \cellcolor{gray2}83.16 & \cellcolor{gray1}98.49 &\cellcolor{gray1}90.30 \\
        Gemini-1.0 & 62.77 & \cellcolor{gray1}92.13 & \cellcolor{gray3}75.21 \\
        Llama-3.1 & \cellcolor{gray3}71.76 & \cellcolor{gray1}97.15 & \cellcolor{gray2}83.26 \\
        \hline
    \end{tabular}
\end{table}

\per has a better generalization capacity than the feature- and pre-training-based methods. 
The AUC of \per for four \llm is: GPT-3.5 (81.64\%), GPT-4o (80.88\%), Gemini-1.0 (81.78\%), and Llama-3.1 (79.55\%), with a range of 2\%. 
The AUC of pre-training-based methods for each LLM is: GPT-3.5 (92.21\%), GPT-4o (90.30\%), Gemini-1.0 (75.21\%), and Llama-3.1 (83.26\%). The max-min difference is over 17\%.
These differences can be explained that pre-training-based methods are trained on the code generated by GPT-3.5, while \per is LLM-independent.
When it comes to feature-based methods, \per outperforms Idialu and Bukhari. 
The AUC of Idialu for each LLM is as follows: GPT-3.5 (91.43\%), GPT-4o (83.16\%), Gemini-1.0 (62.77\%), and Llama-3.1 (71.76\%). The max-min difference is 29\%. The AUC of Bukhari for each LLM is as follows: GPT-3.5 (98.63\%), GPT-4o (98.49\%), Gemini-1.0 (92.13\%), and Llama-3.1 (97.15\%), with a range of 7\%. 
The max-min range of Bukhari is relatively small. 
Bukhari only works for C in which \hac is more different from \lgc than other languages, making the detection method easier to detect \lgc from \hac.

When comparing the generalization capability of methods in \per, we observe that methods with perturbation have better generalization than methods without perturbation. 
Methods with perturbation achieve similar performance on the four LLMs, with the range of average AUC around 3\%. For instance, the average AUC of DetectCodeGPT for the four LLMs is: GPT-3.5 (73.75\%), GPT-4o (73.59\%), Gemini-1.0 (74.90\%), and Llama-3.1 (76.88\%). 
In contrast, methods without perturbation do not exhibit such performance, with some methods fluctuating up to 10\% on different LLMs.
For instance, the average AUC of Rank for the four LLMs is as follows: GPT-3.5 (57.01\%), GPT-4o (54.98\%), Gemini-1.0 (62.68\%), and Llama-3.1 (64.39\%). 
\per without perturbations are based on the absolute perplexity value of the original sample. The absolute perplexity is calculated based on the understanding of the code by the detection model. Therefore, the absolute perplexity varies across code generated by different LLMs, leading to fluctuations in the performance of \per without perturbations.
In contrast, \per with perturbations are based on the perplexity discrepancy between the original and perturbed samples, making them more generalized.

\begin{tcolorbox}[width=\linewidth, boxrule=1pt, sharp corners=all,
  left=2pt, right=2pt, top=2pt, bottom=2pt, colback=gray!8, colframe=black]
  $\textbf{RQ}_\textbf{3}$: \per exhibits a better generalization capability than feature- and pre-training-based methods. The AUC of supervised methods drops significantly when training and testing sets are generated by different \llm.
\end{tcolorbox}

\section{Implications}
\label{SEC:IMP}

This section discusses the characteristics of \lgc and discusses the implications of using \per to detect \lgc.

\subsection{\lgc in Retrospect}

The retrospective analysis of \lgc primarily focuses on its code length and the resulting impact on the evaluation results.

\textbf{\lgc solutions are inherently more concise compared to \hac solutions for identical tasks.} 
The \lgc solutions and \hac solutions in our dataset are used to solve the same programming puzzles. However, our dataset has 4,135 \hac solutions and 8,725 \lgc solutions with fewer than 30 LOC. This indicates that \lgc solutions are inherently more concise than \hac solutions. The conciseness of \lgc solutions can be explained by three main factors: (1) LLMs are familiar with built-in functions and utility libraries, and prefer using them to simplify implementations, (2) LLMs are trained on large-scale code corpora, and tend to imitate standard and concise coding styles, and (3) LLMs prefer minimal implementations rather than verbose or explanatory styles due to the limitation of tokens.

\textbf{The conciseness of \lgc has a negative impact on the detection accuracy of \per.}
\per is unsuitable for detecting short code snippets because of its detection mechanism. Since the first few tokens of a code snippet typically have relatively high perplexity due to limited contexts, short solutions are more influenced by this effect than long solutions and therefore tend to have higher perplexity. As a result, \per may misclassify short \lgc solutions as \hac solutions due to their relatively high perplexity. The results of $\text{RQ}_{1.3}$ show that \per does not perform well on small-scale code with an AUC of only 69.75\%. 
As \per is unsuitable for detecting short code snippets, the conciseness of \lgc solutions increases the difficulty of detection and reduces the detection accuracy of \per.

To mitigate the impact of code length difference on evaluation results, we evaluate the detection methods from the aspect of code scale in RQ3. Specifically, we evaluate the detection methods on three different code scales, where \lgc and \hac solutions within each scale have comparable lengths. In other evaluation aspects, we do not apply any intervention to limit the code length because the differences in code length arise from the inherent characteristics of code written by LLMs and humans, rather than from our designs.

\subsection{For Researchers}
Our implications for researchers include both the pros and cons of \per.
\smallskip

\textbf{\per is unsuitable for high-level programming languages.} 
\per performs well in low-level languages such as C/C++ (91\%) but poorly in high-level languages such as Python (64\%) and Ruby (73\%), which implies that \per has limited capability for high-level languages.
High-level languages encourage a concise and clear code style. High-level languages such as Python provide many built-in operations for basic elements, \eg \emph{list} and \emph{dictionary}; low-level languages have to manually manage \emph{array}, \emph{pointer}, etc. The programming patterns of high-level languages can be more easily learned by a code generator than those of low-level languages.
Moreover, high-level languages have more available third-party libraries than low-level languages. Developers are used to leveraging third-party libraries in their daily development to simplify implementation. Since both humans and \llm use third-party libraries in generating code, detecting \lgc from \hac using \per in high-level languages is challenging. 

\textbf{\per is unsuitable for short code solutions.}
\per achieves an AUC of 87.81\% for code solutions with more than 50 LOC, but its performance drops to 69.75\% for code solutions with less than 20 LOC. This result implies that \per is ineffective for short solutions.
\per is based on the probability that an LLM assigns to the next token being the ground-truth token based on previous tokens. 
Therefore, the perplexity of the first few tokens of code snippets (\ie a collection of tokens) is unstable due to the lack of contexts. In this case, the perplexity of the first few tokens is relatively high. 
When the length of code snippets is short, the high perplexity of the first few tokens significantly affects the average perplexity of the entire code snippets. In this case, the perplexities of \lgc and \hac are both high, which makes \per confused. 

\textbf{\per is unsuitable for just-in-time detection.}
NPR and DetectCodeGPT are the best-performing methods in \per. The detection speeds of NPR and DetectCodeGPT are all around 10 seconds per sample, while the detection speed of the pre-training-based method is less than 0.02 seconds per sample. The difference implies that \per is not suitable for just-in-time detection. 
This can be explained that \per uses an LLM to calculate the perplexity of each token. The efficiency of \llm in calculating perplexity for each token is lower than that of pre-trained models in calculating a probability for an entire code snippet.
Moreover, NPR and DetectCodeGPT are perturbation-based methods. Such methods calculate the perplexity for one original sample and a collection of perturbed samples. As the number of perturbed samples increases, perturbation-based methods become time-consuming.

Although our investigation indicates that \per has several shortcomings, it exhibits advantages in the following aspects.

\textbf{\per exhibits the overall good generalization capability.} 
\per with perturbation achieves stable performance in detecting \lgc across four \llm, with AUC variations of around 2\%. This implies that \per has a good generalization capability when detecting \lgc generated by different \llm. However, \per without perturbation does not achieve such a performance, \eg the AUC of LRR varies by more than 10\% between LLMs. This highlights the importance of incorporating perturbation when using \per to ensure the generalization capability.
The detection mechanism of \per is similar to \llm for code generation. Therefore, \per remains useful provided that \llm are based on the auto-regressive generation mechanism (\ie generating a token based on tokens generated before).
The performances of supervised methods are not well when the code snippets for training and testing are generated by different \llm. The results indicate that supervised methods primarily learn the coding style of LLMs that generate training samples rather than the intrinsic differences between \lgc and \hac. Considering new powerful \llm are emerging, \eg DeepSeek, \per is the most suitable method to address this issue.

\textbf{\per performs well in detecting \lgc in C/C++.} 
\per achieves an AUC of 90.40\% in C and 92.10\% in C++, respectively. This performance is impressive given that \per is a heuristic method rather than a supervised method. These results demonstrate that \per is suitable for detecting \lgc in C/C++. 
C/C++ are low-level programming languages with fewer available third-party libraries than high-level languages such as Python. Therefore, developers implement basic functionality manually. 
Additionally, C/C++ emphasizes more manual control than high-level languages. Therefore, developers handle details like variable types and dynamic memory. 
These characteristics make C/C++ code diverse due to personal coding styles and hard for LLMs to imitate.
\per performs not as well as the pre-training method in C/C++ \lgc of GPT-3.5. However, the pre-training method has poor generalization capability since it is supervised, with an AUC dropping from 92. 21\% on GPT-3.5 to 75.21\% on Gemini-1.0.
In contrast, \per demonstrates the best generalization with the AUC around 80\% on all four LLMs, indicating that \per is LLM-independent. Therefore, considering both performance and generalization, we recommend using \per to detect \lgc in C/C++.

\textbf{\per exhibits the good interpretability.}
\per simulates the code generation mechanisms of \llm to calculate the perplexity of code snippets. Therefore, the perplexity value is interpretable. A code snippet with a high perplexity indicates that it is disjointed, which might not be \lgc. The threshold of perplexity for detection can be adjusted according to the need.
Moreover, the calculation of perplexity is performed token by token, allowing it to be precise for the line-level or even token-level detection. This enables \per to perform fine-grained detection. Fine-grained detection does not mean detecting short code. It refers to detecting code snippets in which only a few lines are generated by LLMs. The detection mechanism makes \per applicable to fine-grained detection. For instance, if most of a code snippet is written by a developer and only a few lines are generated by the LLM, \per can calculate the perplexity line by line and detect the lines with low perplexity as \lgc.
The mechanisms of \per help reviewers understand detection results and provide a foundation for detecting intricate cases where \lgc and \hac are mixed. 
The mechanisms of supervised methods are complex and hard to interpret. They only output a prediction score or binary result, making developers difficult to understand why the code is detected as \lgc and which part is \lgc.  

In short, there is a trade-off between detection accuracy, speed, and generalization capability. \per is not the most effective in terms of accuracy or efficiency, but it offers the best generalization capability. Unlike supervised methods that learn the characteristics of a specific \lgc (\eg GPT 3.5-generated code), \per seeks to understand the intrinsic differences between \lgc and \hac, making it robust when applied to code generated by unknown LLMs. Although \per is LLM-independent and achieves similar performance across different sources, it has suboptimal accuracy and slower detection speed than feature-based and pre-training-based methods. The best-performing methods of \per are based on perturbation, which introduces additional computational overhead and slows the detection. Therefore, \per is not suitable for just-in-time detection.
Separately, \per can be naturally extended to fine-grained detection (\eg line level) when \lgc and \hac are mixed, whereas other methods provide only a binary classification result (or a prediction score) for the entire code snippet. 

Based on the experimental results and the analysis of \per's detection mechanism, we provide the following directions to improve the detection accuracy and speed of \per.

\textbf{Effective perplexity calculation.} 
The current detection LLMs used for perplexity calculation have only a few billion parameters, while LLMs used for code generation in practice typically have hundreds of billions of parameters.
The large difference in parameters results in significant differences in code understanding and generation, as evidenced by the considerable differences in code solutions generated for the same task.
The discrepancy in code understanding and generation between the detection LLM and the generation LLM is a significant reason for the limited effectiveness of \per. 
To address this problem, first, we recommend using larger LLMs to calculate perplexity, which reduces the differences in perplexity understanding between detection LLMs and generation LLMs, but also increases the computational cost. 
Second, we recommend distilling the domain knowledge of LLMs with large parameters into LLMs with small parameters. For instance, we can first collect the code solution generated by LLMs with large parameters, \eg GPT-4o, then utilize fine-tuning or reinforcement learning techniques to train an LLM with small parameters, \eg CodeLlama-7B, in the domain of code generation, and use the distilled LLMs to calculate the perplexity to detect \lgc.
By aligning code understanding between detection and generation LLMs, the effectiveness of \per can be improved.

\textbf{Optimized perturbation method.} 
The perturbation method based on code completion is highly inefficient due to dozens of code masking and completion. Therefore, reducing the number of perturbations is necessary.  
For instance, instead of randomly masking and completing code lines dozens of times, we can first calculate the perplexity of each line in the original code snippet, and then apply only a few perturbations to the lines with high perplexity. This measure not only improves efficiency several-fold but also amplifies the perplexity discrepancies before and after perturbation, which improves the detection effectiveness.
Compared with code completion, adding spaces and empty lines to code snippets is more efficient. However, these perturbation methods are not effective for all types of code, \eg Python.
For indentation-sensitive languages like Python, inserting spaces can cause syntax errors, making the code difficult to understand, \ie leading to high perplexity for both \lgc and \hac.
Therefore, we can further optimize perturbation techniques by considering the characteristics of programming languages. For instance, we can first identify the programming languages of the detected code. If the code is identified as Python, we can apply more targeted perturbation methods, such as replacing variable names. 
The performance of \per can be improved by implementing optimized perturbation methods.

\textbf{Improved perplexity utilization.} 
The current utilization of perplexity is insufficient. Representing the perplexity of a code snippet by the average perplexity of each token leads to significant information loss. Two snippets with the same average perplexity could differ in that one has an overall higher perplexity, while the other has a lower overall perplexity but very high perplexity in certain tokens.
The perplexity of a code snippet forms a continuous distribution based on the perplexity of each token in sequence. Instead of calculating the average perplexity of each token, we need to better leverage the entire perplexity sequence. 
For example, we can count the occurrences where the perplexity exceeds a certain threshold, referred to as surpassing occurrences. \hac tends to be more random and therefore is likely to have more very high perplexity on certain tokens compared to \lgc. 
For another example, we can focus on analyzing changes in the perplexity sequence at the perturbed lines before and after the perturbation. This method amplifies the effect of perturbation, preventing it from being neutralized by the average. The larger the change, the more likely it is \lgc, as \lgc is more sensitive to perturbations~\cite{shi2024between}.
Additionally, we can use the perplexity sequence as the initial input for a supervised classifier. Combining the perplexity sequence with supervised learning can improve detection performance compared with \per (heuristic methods).

We propose the following recommendations for the use of \per.

\begin{itemize}
\item \textbf{Detection method selection: }
    \begin{itemize}
        \item We recommend using DetectCodeGPT or NPR for detection under normal circumstances, as they show the best detection accuracy, acceptable detection speed, and good generalization capability. DetectCodeGPT and NPR perform best among \per, achieving an AUC of around 80\% for different LLMs. Moreover, DetectCodeGPT and NPR demonstrate an acceptable detection speed, detecting a sample in just a few seconds.
        \smallskip
        
        \item We recommend using Log-Rank for detecting large volumes of samples or for just-in-time detection, as it offers acceptable effectiveness and a fast detection speed. Although \per with perturbation offers better detection performance, it takes several tens of seconds to detect a single sample. Log-Rank performs best among \per without perturbation, with an AUC close to 70\% in most cases, while taking only a fraction of a second to detect a sample (50 times faster than DetectCodeGPT/NPR).
    \end{itemize}
    \smallskip

\item \textbf{Application scenario of \per: }
\begin{itemize}
    \item We recommend applying \per to detect \lgc in C/C++ or with a large scale. For the programming language, \per exhibits good performance in C/C++, with an AUC above 90\%, while performing suboptimal in other languages like Python, where it only reaches an AUC of 63\%. For the code scale, \per exhibits good performance in code solutions with a large scale, achieving an AUC of 87\%, while it has limited effectiveness in code solutions with a small or middle scale.
    \smallskip
    
    \item We recommend applying \per when the generation models of \lgc are unknown, as \per has the best generalization capability. \per achieves an AUC of around 80\% regardless of LLMs. However, supervised methods learn the coding styles of LLMs that generate training samples, leading to significant performance fluctuations across different LLMs. Given the continuous emergence of new LLMs and the availability of numerous LLMs for code generation, it is difficult to know whether the training set includes samples generated by the model that generated the code to be detected.    
    \smallskip
    
    \item We recommend applying \per for fine-grained detection since it can be applied to line-level detection. When a code snippet contains both \hac and \lgc, feature- and pre-training-based methods can only assign a single predictive score to the entire snippet and can not make accurate distinctions. In contrast, \per calculates the perplexity token by token, allowing it to assess the perplexity of each line individually.
\end{itemize}
\end{itemize}

\subsection{For Tool Development and Usages}
When detecting \lgc, it is recommended to first detect whether it is \lgc and whether it is entirely or partially generated by \llm, then detect which parts (lines or blocks) are generated by \llm. 

The first step can be formulated as a classification problem. However, the experimental results show that none of the three detection methods (classifiers) are effective. \per only performs well for detecting code in C/C++ (from the perspective of language), and with a large scale (from the perspective of scale). The feature- and pre-training-based methods have high effectiveness but poor generalization capability. Overall, they are all acceptable with an AUC of more than 70\%. An ensemble learning approach can fully use each separate method's advantages while overcoming its disadvantages to some extent.
Taking detecting C code as an example, we recommend using DetectCodeGPT (\per), Bukhari (feature-based method), and GPTSniffer (pre-training-based) to develop an ensemble detection tool. By employing an ensemble learning strategy such as MoE (\underline{M}ixture \underline{O}f \underline{E}xperts), the detection tool can automatically learn which method(s) to prioritize in different contexts.

The second step is fine-grained detection, \ie identifying which lines are generated by \llm. Having understood that the code is generated by \llm, the second step is to first calculate the perplexity of each token in the code and then calculate the average perplexity for each line. These lines with low perplexity are detected as \lgc. We recommend highlighting \lgc and \hac in different colors to remind developers.

The detection tool can be applied in but not limited to following scenarios. 

\textbf{Software supply chain management.} \lgc has become a new third-party asset in software supply chain development~\cite{bukhari2023distinguishing}. The widespread use of \lgc raises concerns regarding code provenance, accountability, and the potential for introducing bugs or vulnerabilities into software. To address these challenges, the detection tool provides a solution for managing \lgc. By detecting the proportion of \lgc in third-party libraries, enterprises can assess the reliability of these libraries. By detecting \lgc in internal code and centrally managing it as a third-party asset, enterprises can verify compliance with licensing requirements to reduce violations. When there is a problem with the code generated by a certain LLM, it can be quickly located and measures can be taken. 

\textbf{Code review.} Since \lgc is more prone to include defects~\cite{perry2024ccsinsecure, sandoval2023lost}, particular attention is required during the code review process. \lgc can lead to unintended risks due to the inherent limitations of LLMs, such as the inclusion of insecure code in the training set~\cite{asare2023github}. Therefore, it is crucial for reviewers to be aware of the potential risks associated with \lgc and to perform extra scrutiny. The detection tool serves as a valuable aid in this process by automatically detecting and highlighting \lgc, ensuring that reviewers are reminded to apply additional checks. These checks may include not only the detection of unique vulnerability patterns targeting \lgc, but also plagiarism detection to ensure that \lgc is compliant with licensing regulations.
   
\textbf{Productivity assessment.} Many developers use LLMs to assist with programming, which has led to a shift in how developer productivity is measured~\cite{shi2024between}. Traditional metrics, such as LOC, have become less applicable because a significant portion of the submitted code is \lgc. LOC-based assessments treat a line of \lgc and \hac as equivalent, resulting in biased evaluations of developers' actual contributions. To mitigate this challenge, the detection tool helps detect \lgc in the submitted code, enabling enterprises to apply more detailed and approximate criteria for assessment. After distinguishing \lgc and \hac, custom evaluations can be performed for each, allowing for a fairer assessment of developer productivity.    
    
\textbf{Originality assessment.} \lgc is sometimes not considered original~\cite{nguyen2024gptsniffer}, which raises a concern in contexts that demand independent problem-solving, intellectual property protection, or academic integrity. The detection tool is useful in such situations by automatically determining whether the submitted code is mainly written by humans or generated by LLMs. It can be applied in various scenarios, including but not limited to software competitions, where independent problem-solving skills are valued, and student assignments, where academic honesty and plagiarism prevention are essential. Using this tool, organizations can ensure that submissions meet the necessary standards of originality.

\section{Threat to Validity}
\label{SEC:TTV}
In this section, we discuss the possible threats to validity and our efforts to mitigate their effects. 

\textbf{Internal Validity: }
The functions in our dataset are designed to solve specific programming puzzles and are mainly self-contained. This may differ from functions in real-world software repositories, where many customized functions exist and cross-file function calls are common. Although we can extract \hac from repositories (\eg GitHub), generating accurate task descriptions from \hac is difficult without domain knowledge. Without task descriptions, LLMs can not generate precise \lgc solutions.
Using \hac and \lgc which solve different problems can introduce bias. For example, using \hac that addresses real-world tasks and \lgc that solves programming puzzles may result in \lgc being less complex than \hac, causing bias in the evaluation results. Therefore, we build our dataset based on CodeNet~\cite{puri2021codenet}, which offers a clear task description for each problem and has been widely used in prior work~\cite{xu2024detecting, dinh2024large, pan2024lost}.

\textbf{Construct Validity: }
We excluded trivial problems in the dataset as they rarely occur in real-world scenarios. According to ~\citet{sheard2013difficult}, there is a strong correlation between code length and problem difficulty. Therefore, we excluded programming problems whose human solutions are all less than 30 LOC to filter out trivial problems. 
We adopted the thresholds used in prior work--10~\cite{xu2024detecting} and 100~\cite{nguyen2024gptsniffer}--to determine the initial range for our threshold setting. Within this range, we tested the number of filtered problems under different thresholds at intervals of 10. When the threshold exceeds 50, the number of filtered problems increases rapidly, and this growth becomes steep when the threshold exceeds 80. Excessive filtering introduces risks of excluding non-trivial problems. Therefore, to ensure the generalization of threshold setting, we adopt a safe value of 30. We manually checked the problems filtered by the threshold of 30. All of them are trivial problems or problems without valid submissions.

\textbf{External Validity: }
The models used for perplexity calculation range from 1 billion to 7 billion parameters. We examine the statistical correlation between the model size and the evaluation results using Spearman’s rank correlation~\cite{spearman1961proof}, which has been widely used in previous work~\cite{niazi2015comparative,hijazi2022quality,lu2012ability,zhou2009examining}. The resulting p-values are all above 0.05, indicating no statistically significant correlation. This can be explained by the fact that, although the models for perplexity calculation differ in parameter size, they are all much smaller than the code-generating models (such as 175 billion). As a result, their code understanding capabilities are all significantly weaker than those of generation models, resulting in no statistically significant correlation between model size and evaluation results.
Small models generally have weaker code understanding capabilities than large models. However, it is impractical for us to locally deploy large generation models (\eg GPT-3.5) due to resource constraints. Therefore, we use the same model as prior work \cite{shi2024between} when calculating perplexity.

\section{Conclusion}
\label{SEC:CON}
\lgc has become increasingly common in daily software development; however, its quality and security remain challenging. 
Moreover, \lgc is sometimes mixed with \hac.
Human-LLM code authorship attribution improves the transparency of software development and increases the trust of third-party software in the supply chain. 
In this article, we conduct a large-scale investigation of \per, the state-of-the-art method for detecting \lgc from \hac. 
Toward realistic evaluations, we use 11,664 \hac snippets and 13,164 \lgc snippets to evaluate \per from three perspectives: detection accuracy, detection speed, and generalization capability. 
The experimental results show that \per is less effective than other methods. \per only performs well in C/C++ from the perspective of language and in a large scale from the perspective of scale. 
Additionally, \per has a slower detection speed than other methods. 
However, \per demonstrates strong generalization capability, making it robust in scenarios where the LLM source is unknown. It also has good interpretability and can be naturally applied to fine-grained detection.
In summary, there is a trade-off between detection accuracy, speed, and generalization capability: \per achieves similar performance across LLMs but has relatively low accuracy and slow speed. While supervised methods achieve higher accuracy and faster speed, their performance varies when the LLMs for code generation are unknown.
In the future, we will continually improve \per to detect \lgc from \hac and evaluate it in industry.

\section*{Data Availability}
The data and materials of this article are available at \url{https://figshare.com/s/10035be3b64e30e6e779}.

\section*{Acknowledgments}
This work is supported by the Key Research and Development Program of Jiangsu Province (No.BE2021002-2), the Natural Science Foundation of Jiangsu Province (No.BK20241195), the Open Project of Key Laboratory of Industry and Information Technology Ministry for Software Fusion Application and Testing Verification, the National Natural Science Foundation of China (No.62302210, No.62202219), and the Innovation Projects and Overseas Open Projects of State Key Laboratory for Novel Software Technology (Nanjing University) (ZZKT2025A12, ZZKT2025B18, ZZKT2025B20, ZZKT2025B22, KFKT2025A17, KFKT2025A19, KFKT2025A20, KFKT2024A02, KFKT2024A13, KFKT2024A14, KFKT2023A09, KFKT2023A10).
Alberto Bacchelli gratefully acknowledges the support of the Swiss National Science Foundation through the SNF Project 200021\_197227.

\bibliographystyle{ACM-Reference-Format}
\bibliography{authorship}

\end{document}